\documentclass[12pt]{article}
\pdfoutput=1
\usepackage[a4paper]{geometry}
\usepackage{color}
\usepackage{float}
\usepackage[utf8]{inputenc}
\usepackage{a4wide}
\usepackage{amsmath,amssymb,amstext,amsthm,amssymb}
\usepackage{amsfonts}
\usepackage{framed}
\usepackage{graphicx}
\usepackage{bbold}
\usepackage{bbm}
\usepackage{slashed} %%feynman slash
\usepackage{tensor} %% tensor indizes
\usepackage{braket} %% bra ket
\usepackage{amstext}
\usepackage{array}
\usepackage{graphicx}
\usepackage{setspace}
\usepackage{hyperref}
\usepackage{overpic}
\usepackage{bbm}
\usepackage{cite}
\usepackage{lipsum}
\usepackage{tikz}
\usepackage{xcolor}

\newcommand{\be}{\begin{equation}}
\newcommand{\ee}{\end{equation}}
\newcommand{\bea}{\begin{eqnarray}}
\newcommand{\eea}{\end{eqnarray}}

\begin{document}

\title{Flat Monodromies and a Moduli Space Size Conjecture}

\pagenumbering{gobble}
\begin{center}
{\huge \textbf {Flat Monodromies and a\\[.3cm] Moduli Space Size Conjecture} } \\ 
\vspace{1.5cm}
{\large Arthur Hebecker$^1$, Philipp Henkenjohann$^1$ and Lukas T. Witkowski$^2$}\\
\vspace{0.5cm}
\textit{$^1$ Institute for Theoretical Physics, University of Heidelberg, \\
Philosophenweg 19, 69120 Heidelberg, Germany\\
\vspace{0.3cm}
$^2$ APC, Universit\'e Paris 7, CNRS/IN2P3, CEA/IRFU, Obs.~de Paris,
Sorbonne Paris Cit\'e, B\^atiment Condorcet, F-75205, Paris Cedex 13, France
(UMR du CNRS 7164)}\\
\vspace{1cm}
\textbf{Abstract}\\
\end{center}
\vspace{0.5cm}

We investigate how super-Planckian axions can arise when type IIB 3-form flux is used to restrict a two-axion field space to a one-dimensional winding trajectory. If one does not attempt to address notoriously complicated issues like K\"ahler moduli stabilization, SUSY-breaking and inflation, this can be done very explicitly. We show that the presence of flux generates flat monodromies in the moduli space which we therefore call `Monodromic Moduli Space'. While we do indeed find long axionic trajectories, these are non-geodesic. Moreover, the length of geodesics remains highly constrained, in spite of the (finite) monodromy group introduced by the flux. We attempt to formulate this in terms of a `Moduli Space Size Conjecture'. Interesting mathematical structures arise in that the relevant spaces turn out to be fundamental domains of congruence subgroups of the modular group. In addition, new perspectives on inflation in string theory emerge. 

\vspace*{10ex}
\noindent August 22, 2017
\newpage
\pagenumbering{arabic}

\section{Introduction}

What properties does a low energy effective field theory have to exhibit to possess a UV completion in the form of a theory of quantum gravity? This is an interesting open problem in theoretical physics whose resolution will have important consequences for cosmology and particle physics \cite{0509212, 0601001, 0605264}. A specific question in this context is, whether quantum gravity sets a limit to the size of field spaces of (pseudo-)scalar fields. This has been studied extensively for axionic fields, i.e.~(pseudo-)scalars with a shift-symmetry. The reason is that such fields may be relevant for 
inflation \cite{Freese:1990rb,Silverstein:2008sg,0808.0706} and cosmological solutions to the hierarchy problem \cite{1504.07551}. 

Many independent approaches to the above set of questions indicate that super-Planckian axionic field ranges are problematic. For one, the Weak Gravity Conjecture (WGC) \cite{0601001} casts doubt on the existence of super-Planckian axions \cite{1409.5793, 1412.3457, 1503.00795, Brown:2015iha,Bachlechner:2015qja,1506.03447,Heidenreich:2015nta}. The electric WGC for axions censors super-Planckian excursions in the field space of one or many axions on the basis of an instanton-induced potential. A naive generalization of the magnetic version of the WGC to axions suggests a hard bound on the axion period at the Planck scale; however, recent more detailed analyses do not provide a definite answer \cite{1701.05572, 1701.06553}.  

Furthermore, one of the conjectures of \cite{0509212,0605264} states that large excursions in field space necessarily lead to effective 4d theories with exponentially small cutoff. We will follow \cite{1610.00010} in calling this the Swampland Conjecture. While it is primarily about non-compact directions in moduli space, one may wonder whether generalizations to other field trajectories exist \cite{1602.06517, 1610.00010, 1703.05776, 1705.04328}, e.g. to those constructed in $F$-term axion monodromy inflation \cite{1404.3040,1404.3542,1404.3711}.

All these findings are consistent with the early observations that axions with a super-Planckian period cannot be straightforwardly obtained from string theory compactifications \cite{0303252, 0605206} (see also \cite{1412.1093}). In addition, entropy bounds were employed in \cite{1203.5476} to argue against super-Planckian axions (see however \cite{Kaloper:2015jcz}). Gravitational instantons may also affect axions \cite{1503.03886, 1607.06814, Alonso:2017avz} and in particular spoil super-Planckian axion field ranges, but a definite answer is elusive without a better understanding of quantum gravity. Also, it is probably fair to say that, quite generally, the mechanism behind the WGC and a possible related censorship of super-Planckian axion field spaces remains obscure.\footnote{
See 
however \cite{Cottrell:2016bty, Hebecker:2017uix, 1705.06287, 1706.08257} for recent work on deriving the WGC from fundamental principles.
}

Given this complicated state of affairs, it is legitimate to take a more optimistic point of view: Why should it not be possible after all to find a string model in which an appropriate scalar potential on top of a small moduli space forces the axion onto a long winding trajectory \cite{Kim:2004rp}? Indeed, a particularly simple implementation of this general idea which uses type IIB fluxes and the gauging mechanism of \cite{Dvali:2005an} has already been described in \cite{1503.07912} under the name of `Winding Inflation'. In this way, one would at least have an `effective' long-range axion (see \cite{Saraswat:2016eaz} for related considerations in the 1-form context). However, also this optimistic example-based attitude has remained unconvincing because of the complications of any realistic string construction. 

In this work, we attempt to resolve the issue by constructing super-Planckian axion field spaces in a very simple stringy setting, which allows for explicit calculations. We are not interested in inflation or any other phenomenological application, which allows us to avoid the problems of realistic string constructions. In addition, we are not prone to the (possibly model-dependent) backreaction effects which underlie the bounds obtained in \cite{1602.06517, 1703.05776}. This increases the chance that any bounds we find have a generic quantum gravity origin. 

Thus, our focus are supersymmetric, \emph{flat} axionic directions such that backreaction plays no role. This is close in spirit to the approach taken in \cite{1601.00647}. Here we choose to work with type IIB string theory compactified on a toroidal orientifold with supersymmetric 3-form flux. Such a flux generically reduces the dimension of moduli space. It can also introduce a monodromy (with finite but possibly large monodromy group) in the remaining flat directions.\footnote{Monodromies also arise in flux compactifications on Calabi-Yau manifolds and have been discussed in the context of moduli dynamics and tunneling in the string landscape, see e.g.~\cite{0612222, 0805.3705, 1011.6588, 1108.1394}. In these works monodromy transformations connect points with different values of the scalar potential or isolated vacua. By contrast, we study monodromy transformations between points on a periodic flat direction, enlarging the periodicity of the latter. Note also that, following the recent literature on inflation, we use the term monodromy for the breaking of a periodicity by flux, not for the large diffeomorphism required to make the original periodicity manifest.} To keep the discussion focussed on the question at hand, we do not address the problem of stabilizing the remaining moduli. Working out the consequences of our flux choice we find that a certain two-dimensional subspace of the full moduli space is enlarged by a factor $N$, where $N$ is a flux number. In this way, to the best of our present understanding, a super-Planckian flat axionic direction emerges.

However, one should be careful about an interpretation of this in the sense of a large field space. The key is the geometry of this space. Indeed, the reason for the extended moduli space is the reduced modular invariance of tori with fluxes as compared to tori without flux. The resulting moduli space is given by a fundamental domain of so-called congruence subgroups of SL$(2,\mathbbm{Z})$. Together with the proper metric, this space is a Riemann surface of a certain genus, with locally hyperbolic geometry, with a number of conical singularities and with singular cusps or throats. The natural way to measure distances between two points in this space is via geodesics. However, the long axionic trajectories advertised above are very far from being  geodesics. Two points on such an axionic trajectory may have an `axionic' distance $\sim N$, with $N$ our potentially large flux number. Yet their geodesic distance is only $\sim \ln(N)$. More generally the geodesic distance between any two points is bounded by an expression of order $\ln(1/\Lambda)$, where $\Lambda$ is the cutoff below which the 4d effective theory is valid.

We try to formalize these findings in terms of two conjectures which are related to but also distinctly different from the well-known Swampland Conjecture and recent variants \cite{0509212,0605264,1610.00010,1705.04328}. Consider the moduli space of a generic 4d field theory with cutoff $\Lambda$. Then we conjecture that the absolute size of the moduli space, as measured by the appropriately defined diameter, scales as $\ln(1/\Lambda)$. Alternatively, we may focus on the full moduli space of a certain string compactification. Pick two points in this moduli space which are connected by a geodesic with length $L$. Then we claim that there exist points on this geodesic at which the lightest KK or winding mode mass is smaller or of the order of $\exp(-\alpha L)$, with $\alpha\sim {\cal O}(1)$.

At first sight all of this might suggest that long and in particular long axionic trajectories are not realizable in 4d effective field theories with high cutoff. However, recall that we have found a long axionic direction. The fact that this direction was not a geodesic may be irrelevant if one is able to construct an appropriate potential that forces the field onto this long trajectory.\footnote{Recently, a model of inflation has been proposed in which the hyperbolic geometry of field space is essential \cite{Brown:2017osf} (see also \cite{Mizuno:2017idt}). It would be interesting to see whether this can be realized in our setting. Such models have also been discussed in \cite{Achucarro:2016fby,Achucarro:2017ing} (see also \cite{Kobayashi:2010fm,Cremonini:2010sv,Cremonini:2010ua,vandeBruck:2014ata,Turzynski:2014tza}). In particular it has been pointed out therein that compatibility with observation may be achieved without stabilizing all scalar fields except for the inflaton itself.} Thus, it appears that the question of large-field inflation requires knowledge beyond the Weak Gravity and Swampland Conjectures.

\section{A monodromic moduli space via fluxes}
\subsection{KNP vs.~winding trajectories from fluxes}

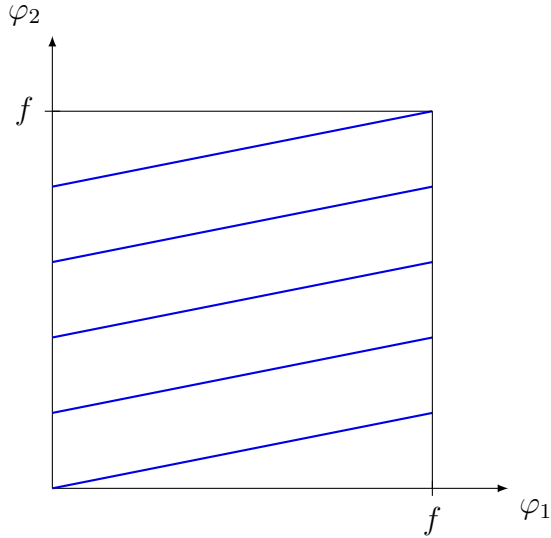
\begin{figure}[t]
\begin{center} 
\begin{tikzpicture}
\draw[-latex] (0,0) -- (6,0) node[anchor=north west] {$\varphi_1$};
\draw[-latex] (0,0) -- (0,6) node[anchor=south east] {$\varphi_2$};
\draw[thick,color=blue!90!black] (0,0) -- (5,1);
\draw[thick,color=blue!90!black] (0,1) -- (5,2);
\draw[thick,color=blue!90!black] (0,2) -- (5,3);
\draw[thick,color=blue!90!black] (0,3) -- (5,4);
\draw[thick,color=blue!90!black] (0,4) -- (5,5);
\draw (5,0) -- (5,5);
\draw (0,5) -- (5,5);
\draw (5,-0.1) node[anchor=north] {$f$} -- (5,0.1);
\draw (-0.1,5) node[anchor=east]{$f$} -- (0.1,5);
\end{tikzpicture}
\caption{Winding flat direction of total length $\sim Nf$ (shown for $N=5$).}
\label{winding} 
\end{center}
\end{figure}

We want to construct a long axionic direction in the moduli space of a supersymmetric compactification of type IIB string theory as a long winding trajectory in a compact field space. This is the Kim-Nilles-Peloso (KNP) mechanism  \cite{Kim:2004rp}, but in our case the winding trajectory will arise due to 3-form fluxes rather than the instanton potential employed in \cite{Kim:2004rp}. The idea is as follows. Consider a theory with two axions $\varphi_1$ and $\varphi_2$ with small and, for simplicity, equal periodicity given by the axion decay constant $f$. Even though this field space is small one can generate a long trajectory by having a potential for the axions that has a minimum at $\varphi_1=N\varphi_2$ for a large integer $N$. Now, the remaining flat direction has a periodicity of $\sqrt{N^2+1}f$ which is much larger than the original $f$ for large $N$ (see Fig.~\ref{winding}). This is the KNP-mechanism.

We choose to work in a simple setup of toroidal orientifolds. Thus we take as the compact space $\text{T}^6 / \text{Z}_2 = (\text{T}_1^2 \times \text{T}_2^2 \times \text{T}_3^2) / \text{Z}_2$, i.e.~a factorisable 6-torus subject to a $\text{Z}_2$ identification. By turning on 3-form fluxes on the tori we will show how one can generate a superpotential of the form \cite{0201028,0506179,1405.0283,1703.09729}
\begin{align}
\label{superpot} W = (M\tau_1 - N\tau_2) (\tau - \tau_3) \, ,
\end{align} 
where $\tau= C_0 + \text{i} e^{- \phi}$ is the axio-dilaton, $\tau_i$ with $i=1,2,3$ are the complex structure moduli of the three 2-tori and $M,N$ are integers (flux numbers). For the following analysis it will be useful to label the real and imaginary components of $\tau$ and $\tau_i$ and we hence define
\begin{align} 
\tau &= C_0 + \text{i} e^{- \phi} = c+\text{i}s \, , \\
\tau_i &= \text{Re}\,\tau_i+\text{i}\,\text{Im}\,\tau_i = u_i + \text{i} v_i \, .
\end{align}
Throughout this work we will refer to the real parts $c = \textrm{Re} \, \tau$ and $u_i = \textrm{Re} \, \tau_i$ as `axionic' directions due to their associated shift symmetries.\footnote{In the case of $c$ the shift symmetry arises from the $SL(2, \mathbbm{Z})$ symmetry of type IIB string theory and persists beyond toroidal orientifold compactifications. The shift symmetries in $u_i$ originate from the $SL(2, \mathbbm{Z})$ modular symmetries of the compactification tori. For more general compactifications on Calabi-Yau threefolds, shift symmetries in the complex structure moduli sector are typically broken, but this breaking becomes increasingly weak when approaching large complex structure.}

Without loss of generality we can take $v_i >0$. A minimum of the scalar potential is determined by the conditions $D_IW=0$ and $W=0$, where $I$ runs over all moduli. This corresponds to the supersymmetric vacuum with $\tau_3=\tau$ and $M\tau_1=N \tau_2$. Note that the minimum is not a unique point in field space, as there are several flat directions. First, let us consider only one particular flat direction in the $(u_1, u_2)$ field space, defined by
\begin{align}
\psi \equiv M u_1 - N u_2 = 0
\label{flatdir}
\end{align}
and all other moduli fixed. Our main focus is whether this direction can be long enough such that we can traverse a trans-Planckian distance.

Naively, it may seem that there is no bound to this flat direction. If we increase $u_1$ we simply have to increase $u_2$ accordingly to keep $\psi=0$. Of course, as suggested by Fig.~\ref{winding}, we will return to the same geometrical situation after a certain distance. But it is at first sight not obvious whether the flux configuration on the torus has changed.

To study this in detail, recall $u_1$ and $u_2$ are the real parts of $\tau_1$ and $\tau_2$, which are the complex structure moduli of two tori. Further recall that the complex structure moduli sector exhibits a modular symmetry: All tori whose complex structure moduli are related by an SL$(2, \mathbbm{Z})$ transformation are equivalent. Thus, if we wish to limit ourselves to physically inequivalent configurations, we have to limit the range of $\tau_1$ and $\tau_2$ to the fundamental domain of SL$(2, \mathbbm{Z})$. Accordingly, $u_1$ and $u_2$ are constrained to be in the corresponding fundamental domain.

However, the situation becomes more complicated in the presence of 3-form fluxes. Since these are 3-forms on the tori, a modular transformation on them will also induce a transformation of the fluxes. In the following, we show how this leads to a monodromic, i.e.~enlarged, moduli space and to a long but finite axionic direction.

\subsection{Brief interlude concerning the action of the modular group}
Before we explain how to arrive at a superpotential (\ref{superpot}) and how the moduli space is extended we need to set up some elementary notation concerning SL$(2,\mathbbm{Z})$ and gauge redundancies of tori. Let a torus be defined as the complex plane modded out by some lattice,
\be
\mathbbm{C}/\mbox{span}_{\mathbbm{Z}}(e_y,e_x)\,.
\label{latticevec}
\ee
Coordinates $y\in [0,1)$ and $x\in [0,1)$ are introduced by
\be
z=(y,x)\cdot\left(
\begin{array}{c}
e_y \\ e_x
\end{array}
\right)\,.
\ee
For example, with $e_y=\tau$, $e_x=1$ we have
\be
z=(y,x)\cdot\left(
\begin{array}{c}
\tau \\ 1
\end{array}
\right)=x+\tau y\,.
\ee
More generally, the same torus is described by
\be
z\,=\,(y,x)\,R^{-1}R\left(
\begin{array}{c}
\tau \\ 1
\end{array}
\right)\,=\,e_x'x'+e_y'y'\,\equiv\, e_x'\,(x'+\tau' y')\,,
\ee
with 
\be
R= \left(\begin{array}{cc} a & b \\ c & d \end{array}\right)\in SL(2,\mathbbm{Z})\,\,,\qquad 
\tau'\equiv \frac{e_y'}{e_x'}=\frac{a\tau + b}{c\tau+d}\equiv R(\tau)
\ee
and
\be
\left(\begin{array}{c} y' \\ x' \end{array}\right)
=R^{-1\,\text{T}} \left(\begin{array}{c} y \\ x \end{array}\right)\,. 
\ee
For our following analysis it will be important that, by the above logic, the components of any 1-form 
\be
\omega=\omega_i \text{d}\xi^i\qquad\mbox{with}\qquad \text{d}\xi^i=
\left(\begin{array}{c}
\text{d}y \\ \text{d}x
\end{array}\right)
\ee
transform according to 
\be
\omega_i'=R_i{}^j\omega_j\,.
\ee

\subsection{Flux choice}

Let us briefly describe how we can arrive at a superpotential of the form \eqref{superpot} from flux compactifications in toroidal orientifolds. Here and in the following we will set $(2 \pi)^2 \alpha'=1$. The superpotential is the Gukov-Vafa-Witten superpotential, which can be written as
\begin{align}
W = \int_X \Omega_3 \wedge G_3 \, ,
\end{align}
where 
\begin{align}
\Omega_3 & = \text{d}z_1\wedge\text{d}z_2\wedge\text{d}z_3 \\
\nonumber & = (\text{d}x_1 + \tau_1 \text{d} y_1) \wedge (\text{d}x_2 + \tau_2 \text{d} y_2) \wedge (\text{d}x_3 + \tau_3 \text{d} y_3)\,,\\
\nonumber G_3 & = F_3 - \tau H_3\, ,
\end{align}
and $(y_i,x_i)$ are the coordinates on the $i$th torus. For completeness, let us also record the K\"ahler potential
\begin{align}
\nonumber K &= -\ln \left( -i (\tau -\bar{\tau})\right) - 2 \ln \mathcal{V} - \ln \left( -i \int_X  \Omega_3 \wedge \overline{\Omega}_3 \right) \\
&= -\ln \left( -i (\tau -\bar{\tau})\right) - 2 \ln \mathcal{V} - \ln \left( i (\tau_1 -\bar{\tau}_1)(\tau_2 -\bar{\tau}_2)(\tau_3 -\bar{\tau}_3)\right) \, .
\label{kahlerpot}
\end{align}

The superpotential in \eqref{superpot} then arises for the following choice for the 3-form fluxes:\footnote{Note that odd flux numbers $M$ and $N$ imply the existence of further `exotic' O3 planes \cite{0201028}.}
\bea
F_3&=&(+M\,\text{d}x_1\wedge \text{d}y_2-N\,\text{d}y_1\wedge \text{d}x_2)\wedge \text{d}x_3\, ,
\\
H_3&=&(-M\,\text{d}x_1\wedge \text{d}y_2+N\,\text{d}y_1\wedge \text{d}x_2)\wedge \text{d}y_3\,.
\label{fluxansatz}
\eea
Note that this can also be written more compactly as $F_3=+\mathcal{A}\wedge \text{d}x_3$ and $H_3=-\mathcal{A}\wedge \text{d}y_3$, where we introduced the 2-form $\mathcal{A}$ which is only supported on the first two tori:
\be
\mathcal{A}=A_{ij}\,\text{d}\xi_1^i\wedge \text{d}\xi_2^j\qquad\text{with}\qquad\xi_1^i=
\left(\begin{array}{c} y_1 \\ x_1 \end{array}\right)
\qquad\mbox{and}\qquad \xi_2^i=\left(\begin{array}{c} y_2 \\ x_2 \end{array}\right)\,.
\ee
The essential part of the explicit flux information is encoded in the matrix
\be
A_{ij}=\left(\begin{array}{cc} 0 & -N \\ M & 0 \end{array}\right)\,.
\ee
This flux choice enforces $M\tau_1=N\tau_2$ and $\tau_3=\tau$. We will ignore $\tau$ and $\tau_3$ and focus on the restricted 2-dimensional moduli space resulting from $\tau_1$ and $\tau_2$. It can be parametrized, for example, by $\tau_1$ alone.

There is a constraint on the values of $N$ and $M$ coming from the $D3$ tadpole cancellation condition. It reads
\begin{align}
N_{D3} + \frac{1}{2} \int_X H_3 \wedge F_3 = \frac{1}{4} N_{O3} \, ,
\end{align}
where $N_\text{D3}$ is the number of D3-branes and $N_\text{O3}$ is the number of O3-planes. For the toroidal orientifold $\text{T}^6 / \text{Z}_2$ one finds 64 fixed points corresponding to 64 O3-planes. The flux contribution for our ansatz (\ref{fluxansatz}) can be calculated as $\int_X H_3 \wedge F_3 = 2MN$. We thus arrive at the constraint:
\begin{align}
MN \leq 16 \, ,
\end{align}
where the maximal value of 16 is attained for $N_{D3}=0$. 

\subsection{The monodromic moduli space}
Let us now return to the question of the size of moduli spaces in the presence of flux. Given our superpotential \eqref{superpot} the minimum at $W=0$ exhibits two complex flat directions defined by $(\tau-\tau_3)=0$ and $(M\tau_1-N \tau_2)=0$. Here we will focus on the latter.

As noted before, we can restrict attention to $\tau_1$. Naively, one expects it to take values e.g.~in the canonical fundamental domain. We will immediately see that, in the presence of fluxes, this is not any more true. Consider an arbitrary $\tau_1$ and a flux configuration determined by the matrix $A$. Now, while keeping $A$ fixed, move $\tau_1$ in the upper complex half plane to any other $\tau_1'$ that is related to $\tau_1$ by a modular transformation, i.e.
\be
\tau_1=R_1(\tau_1')=\frac{a\tau_1'+b}{c\tau_1'+d}
\ee
for some $R_1\in\text{SL}(2,\mathbbm{Z})$. Then the $(F_3,H_3)$ fluxes also transform nontrivially due to the transformation properties of the matrix $A_{ij}$:
\be
A_{ij}\to A_{ij}'=(R_1)_i{}^k A_{kj}\,.
\label{fluxtrafo}
\ee
Therefore, although $\tau_1$ and $\tau_1'$ are related by a modular transformation and the corresponding two tori are identical, the whole physical configuration may be different due to different values of the fluxes given by (\ref{fluxtrafo}). However, it is possible that this non-trivial transformation of the fluxes can be undone by a transformation acting on the second index, associated with a modular transformation of the second torus. For this, one must require that an SL$(2,\mathbbm{Z})$ matrix $R_2$ exists such that
\be
A''=R_1 A R_2^T = A\,.
\ee
The condition for this to be possible is that the matrix $A^{-1}R_1^{-1}A$ is in SL$(2,\mathbbm{Z})$,
\be
R_2^\text{T}=A^{-1}R_1^{-1}A=\left(\begin{array}{cc}
a & cN/M \\ bM/N & d \end{array}\right) \in \text{SL}(2,\mathbbm{Z})\,.
\label{redsl}
\ee
Restricting our attention to the case where $M$ and $N$ have no common divisors, $b$ must be a multiple of $N$ and $c$ a multiple of $M$.

An important consistency check is to verify that, after performing the transformations above, we still satisfy the vacuum condition $M\tau_1'=N\tau_2'$. Indeed, one easily calculates
\begin{equation}
N\tau_2'=N\frac{a\tau_2+bM/N}{cN\tau_2/M+d}=M\tau_1'\, ,
\end{equation}
where we used $M\tau_1=N\tau_2$.

\begin{figure}[h]
\begin{center}
\includegraphics[width=0.4\textwidth]{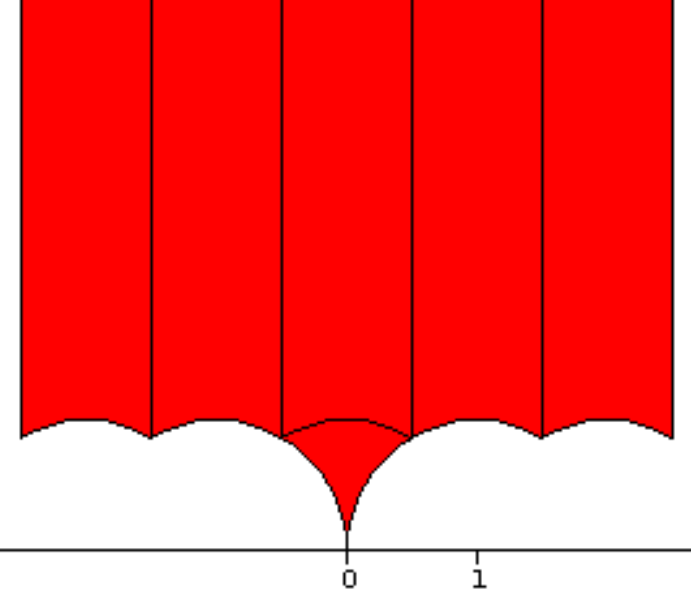}
\end{center}
\caption{A fundamental domain of the congruence subgroup $\Gamma^0(5)$ as a subset of the upper complex half plane is shown. The central strip without the `triangle' touching the real axis corresponds to the standard fundamental domain of the complex structure modulus of a torus.}
\label{fundomainn5}
\end{figure}

In the special case of $M=1$, the only restriction on $R_1$ is that $b$ is a multiple of $N$. This means that the `smallest' transformation of $\tau_1$, defining the periodicity of its real part, takes the form
\begin{equation}
R_1=
\begin{pmatrix}
1 & N \\
0 & 1
\end{pmatrix}\,.
\end{equation}
But this is exactly what we expected: The width of the fundamental domain is not unity, e.g.~Re$\,\tau_1\in (-1/2,1/2)$, but has been extended to $N$, such that we can choose e.g.~Re$\,\tau_1\in (-N/2,N/2)$. Fig.~\ref{fundomainn5} shows such an extended fundamental domain for $N=5$, calculated with the program `fundomain' by H.~Verrill \cite{Verrill:2001}. Since this is the only feature of interest for us we set $M=1$ throughout the rest of the paper.

Using the K\"ahler potential (\ref{kahlerpot}) one can determine the metric in moduli space restricted to $\tau_1$ and $\tau_2$:
\begin{equation}
\text{d}s^2=\frac{\text{d}\tau_1\text{d}\overline{\tau_1}}{4(\text{Im}\,\tau_1)^2}+\frac{\text{d}\tau_2\text{d}\overline{\tau_2}}{4(\text{Im}\,\tau_2)^2}\,.
\label{modmetrickae}
\end{equation}
Evaluating this in the vacuum $\tau_1=N \tau_2$ parametrized by $\tau_1$ one finds
\begin{equation}
\text{d}s^2=\frac{\text{d}\tau_1\text{d}\overline{\tau_1}}{2(\text{Im}\,\tau_1)^2} \, .
\label{modmetric1}
\end{equation}
We are now in a position to calculate the length of the flat direction defined in \eqref{flatdir}. Our result is
\begin{align}
\label{eq:naiveresult}
L = \int_{-N/2}^{N/2} \frac{\textrm{d} u_1}{\sqrt{2}\,\text{Im}\,\tau_1} = \frac{N}{\sqrt{2}\,\text{Im}\,\tau_1} \, .
\end{align}
Note that the value of $N$ is bounded by a tadpole constraint such that $N=16$ is the largest allowed value. Saturating this bound and setting Im$\,\tau_1=1$ we find $L = 8\sqrt{2}$ for the length of the flat direction. To summarize, it appears that we have succeeded in generating a super-Planckian flat axionic direction.

\section{Topology and geometry of fundamental domains of congruence subgroups}

\begin{figure}[t]
\begin{center}
\includegraphics[width=0.5\textwidth]{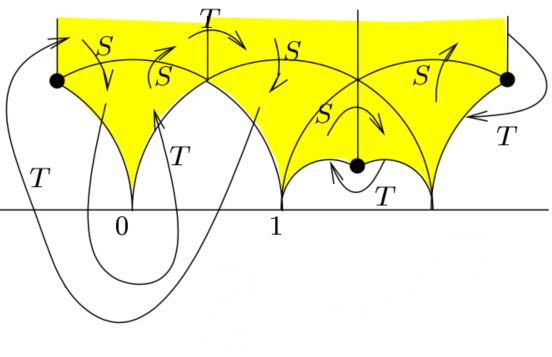}
\end{center}
\caption{The lower part of the fundamental domain of the congruence subgroup $\Gamma^0(7)$ is shown. Appropriate identifications of the boundaries are indicated \cite{
Verrill:2001}.}
\label{identifications}
\end{figure}

The transformations described by Eq.~\eqref{redsl} constitute so-called congruence subgroups of SL$(2,\mathbbm{Z})$. We have already shown the fundamental domain of such a subgroup for the case $M=1$ and $N=5$, denoted by $\Gamma^0(5)$ in Fig.~\ref{fundomainn5}. We can explicitly see the enlarged field space for $\text{Im}\tau_1>1$ in the direction parallel to the real axis. The vertical boundaries on the left and right of the fundamental domain are identified as is the case for the standard fundamental domain of $\text{SL}(2,\mathbbm{Z})$. However, the identifications in the bottom are much more subtle. Fig.~\ref{identifications} shows the lower fundamental domain of $\Gamma^0(7)$ with the appropriate identifications indicated \cite{
Verrill:2001}.

\begin{figure}[h]
\begin{center}
\includegraphics[width=0.3\textwidth]{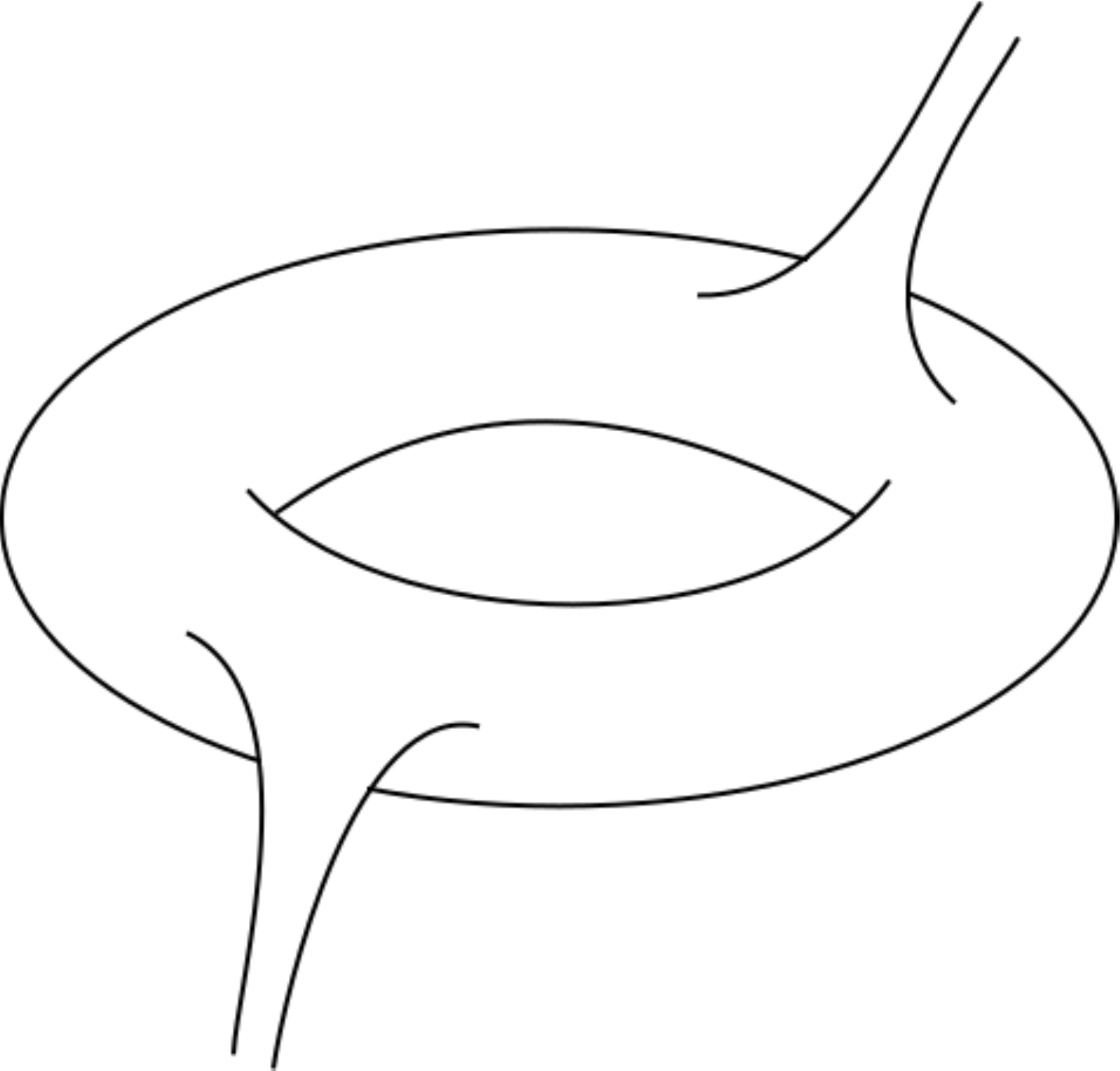}
\end{center}
\caption{A qualitative picture of a fundamental domain of a congruence subgroup as a Riemann surface. The throats correspond to the cusps of the fundamental domain together with the point at infinity.}
\label{riemann}
\end{figure}

\begin{figure}[h]
\begin{center}
\includegraphics[width=0.7\textwidth]{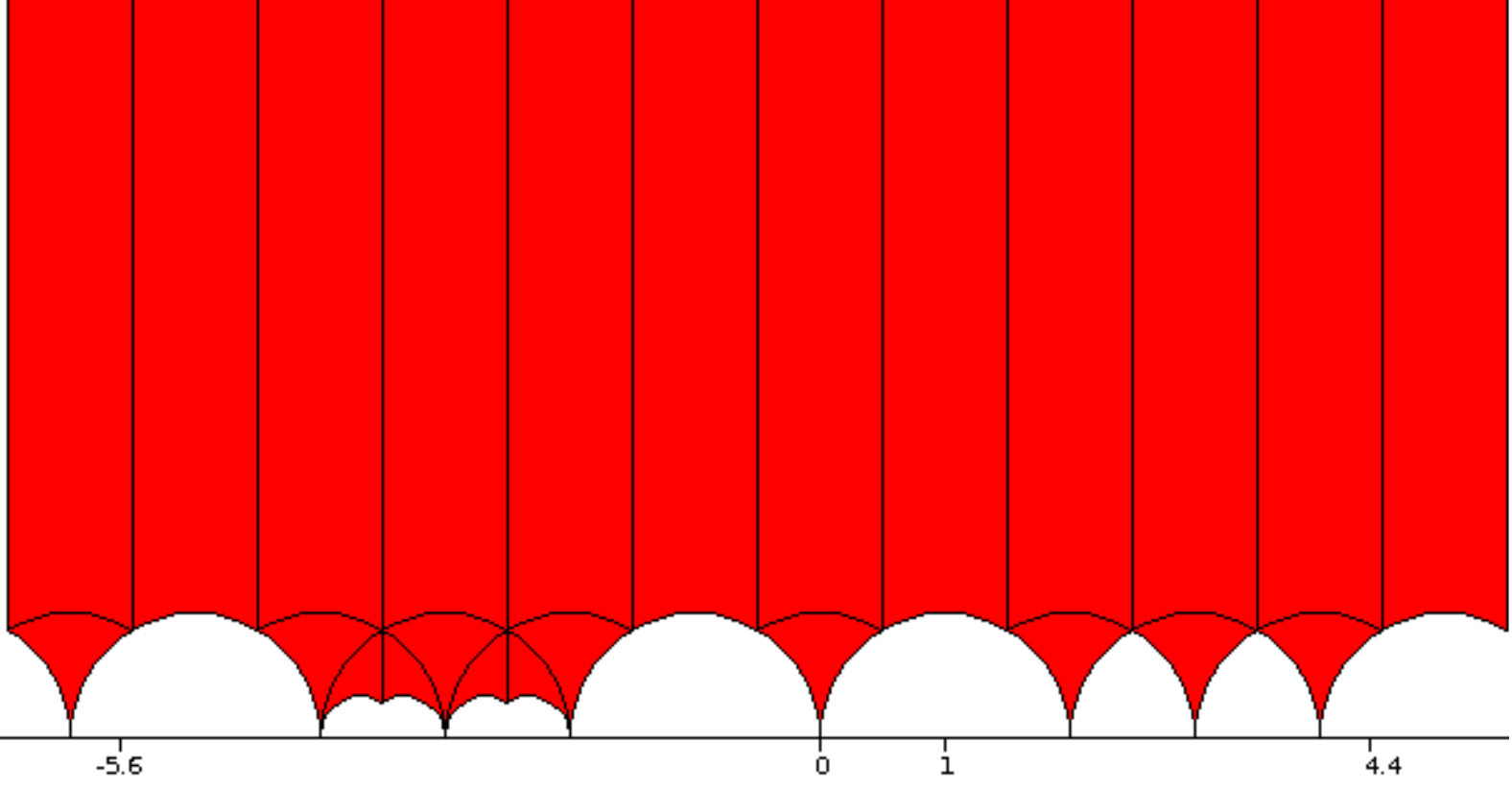}
\end{center}
\caption{A fundamental domain of the congruence subgroup $\Gamma^0(12)$ with several cusps is shown.}
\label{fundomain12}
\end{figure}

Recall the metric on the moduli space of one torus (see e.g.~\eqref{modmetrickae}),
\begin{equation}
\text{d}s^2=\frac{\text{d}u^2+\text{d}v^2}{4v^2}\,,
\label{modmetric}
\end{equation}
where $u$ is identified with the real and $v$ with the imaginary part of the relevant complex structure modulus. This metric is the natural metric on the space of all tori with fixed volume. The upper complex half plane equipped with this metric is the hyperbolic plane. Fundamental domains of SL$(2,\mathbbm{Z})$ and its congruence subgroups can therefore be viewed as subsets of this plane (with appropriate identifications of boundaries). They can have different topologies (non-trivial genus), cusps and conical singularities \cite{Stein:2003}. A qualitative picture of such a Riemann surface is shown in Fig.~\ref{riemann}. The throats in the picture correspond to the cusps in the fundamental domain where it extends to the real axis. Also, the point at infinity in the complex half plane gives rise to such a throat. As one can see in Fig.~\ref{fundomain12} for the congruence subgroup $\Gamma^0(12)$, there may be several of these cusps. The picture also clearly shows the widened fundamental domain, now by a factor 12, compared to the fundamental domain of a torus.

Let us now discuss the potentially long axionic directions corresponding to lines of $\text{Im}\,\tau_1=\text{const.}$ Using the metric (\ref{modmetric}) we see that the length of these lines increases with decreasing  $\text{Im}\,\tau_1$. However, the smallest value of $\text{Im}\,\tau_1$ that allows for a straight unbroken line is $\text{Im}\,\tau_1=1$. This is a direct consequence of the complicated structure of the fundamental domain at $\text{Im}\,\tau_1<1$. We have already calculated the periodicity of this axionic direction to be $N/\sqrt{2}$. In our setting a tadpole condition bounds $N$ by $16$ from above which therefore quantifies the maximal length of these axionic directions. We expect that corresponding lengths in more involved compactification on CY's in the large complex structure limit surpass this significantly.

So far this sounds very encouraging. However, as long as there is no potential for $\tau_1$, straight lines defined by $\text{Im}\,\tau_1=\text{const}$.~are by no means the most natural paths connecting two points on this line. In fact they are not geodesics with respect to the proper metric (\ref{modmetric}) on moduli space, i.e.~there exist shorter paths. It is therefore somewhat arbitrary to declare these non-geodesic paths to be long since one can always generate long paths by means of a detour.

\begin{figure}[h]
\begin{center}
\begin{tikzpicture}
\fill[black!10!white] (-5,0 ) rectangle (5,1);
\draw[-latex] (-5.5,0) -- (5.5,0) node[anchor = north west] {Re$\,\tau_1$};
\draw [-latex] (0,0) -- (0,6) node[anchor = south east] {Im$\,\tau_1$};
\draw (-5,0.1) -- (-5,-0.1) node[anchor=north] {$-\frac{N}{2}$};
\draw (5,0.1) -- (5,-0.1) node[anchor=north] {$\frac{N}{2}$};
\draw (0.1,1) -- (-0.1,1) node[anchor=east] {1};
\draw (0.1,2) -- (-0.1,2) node[anchor=east] {$a$};
\draw (0.1,4.5) -- (-0.1,4.5) node[anchor=east] {$b$};
\draw  (3.5,0) --(3.5,2);
\draw  (3.5,4.5) --(3.5,5.5);
\draw [line width=0.5mm] (3.5,2) --(3.5,4.5);
\filldraw (-5,1) circle (0.05);
\filldraw (5,1) circle (0.05);
\draw (5.1,0) arc (0:11:5.1);
\draw [line width= 0.5mm] (11:5.1) arc (11:150:5.1);
\draw (150:5.1) arc (150:180:5.1);
\draw [dashed](0,0) -- (11.25:5.1);
\draw [dashed](0,0) -- (150:5.1);
\draw (5.5:1.5) node {$\alpha$};
\draw (170:1.5) node {$\pi-\beta$};
\draw (1.7,0) arc (0:11:1.7);
\draw (-2.2,0) arc (180:150:2.2);
\draw [dashed] (0,2) -- (3.5,2);
\draw [dashed] (0,4.5) -- (3.5,4.5);

\end{tikzpicture}
\end{center}
\caption{The two types of geodesics of the hyperbolic plane are shown: a vertical line and a semi-circle. The segments of these of which the length is calculated in the main text are drawn with thick lines. The shaded region at the bottom corresponds to the region where the fundamental domains of $\Gamma^0(N)$ are in general very complicated (see also Figs.~\ref{identifications} and \ref{fundomain12}).}
\label{geodesics}
\end{figure}
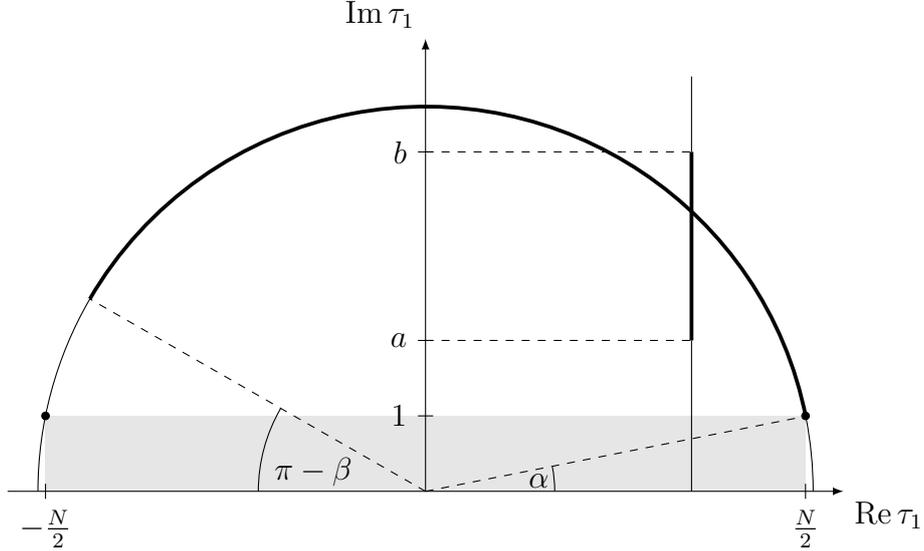

It turns out that geodesics of the hyperbolic plane are given by lines of constant $\text{Re}\,\tau_1$ and arcs of circles with their center on the real axis (see Fig.~\ref{geodesics}). Let us calculate the length of these geodesics. We start with the straight lines of constant real part and consider only a segment of one of these lines starting at $\text{Im}\,\tau_1=a$ and ending at  $\text{Im}\,\tau_1=b$. The length is given by
\begin{equation}
L=\int_a^b\frac{\text{d}y}{2y}=\frac{1}{2}\ln\left(\frac{b}{a}\right)\,.
\label{linelength}
\end{equation}
This is the well-known logarithmic behavior of proper field displacements in moduli space. Now let us calculate the length of an arc of a circle with radius $R$ which starts at a polar angle $\alpha$ and ends at an angle $\beta$. The center of this circle may be located anywhere on the real axis. Parameterizing this path by the polar angle one finds
\begin{equation}
L=\int_\alpha^\beta\text{d}\varphi\frac{R}{2R\sin(\varphi)}=\frac{1}{2}\ln\left(\frac{\tan(\beta/2)}{\tan(\alpha/2)}\right)=\frac{1}{2}\ln\left(\frac{1/\sin(\beta)-1/\tan(\beta)}{1/\sin(\alpha)-1/\tan(\alpha)}\right)\,.
\end{equation}
For a symmetric arc with $\beta=\pi-\alpha$ this can be simplified to
\begin{equation}
L=\frac{1}{2}\ln\left(\frac{1+\cos(\alpha)}{1-\cos(\alpha)}\right)\,.
\label{symmarc}
\end{equation}

Using this formula, we now consider deformations of our long axionic trajectory and determine how short it can become. Indeed, Fig.~\ref{geodesics} shows the long, closed axionic trajectory as a horizontal line connecting the point $-N/2+i$ with the (equivalent) point $N/2+i$. It can be deformed to the arc, also shown in the figure, which again connects this fixed point with itself. For large $N$ and hence small $\alpha$ the result is approximately $L\approx\sqrt{2}\ln(N/2)$, which is clearly much less than our naive expectation in \eqref{eq:naiveresult}, which grew linearly with $N$.\footnote{Compared to (\ref{symmarc}) this expression for $L$ contains an additional factor $\sqrt{2}$ in order to take the contribution from $\tau_2$ to the length into account, see also (\ref{modmetrickae}) and (\ref{modmetric1}). In the following we will tacitly include this factor in expressions for lengths when appropriate.} The upshot is that even if we manage to construct models with large $N$ and hence long axionic directions, we have to be very cautious about the question to which extent these represent large proper distances between points in field space.

\begin{figure}[h]
\begin{center}
\includegraphics[width=0.5\textwidth]{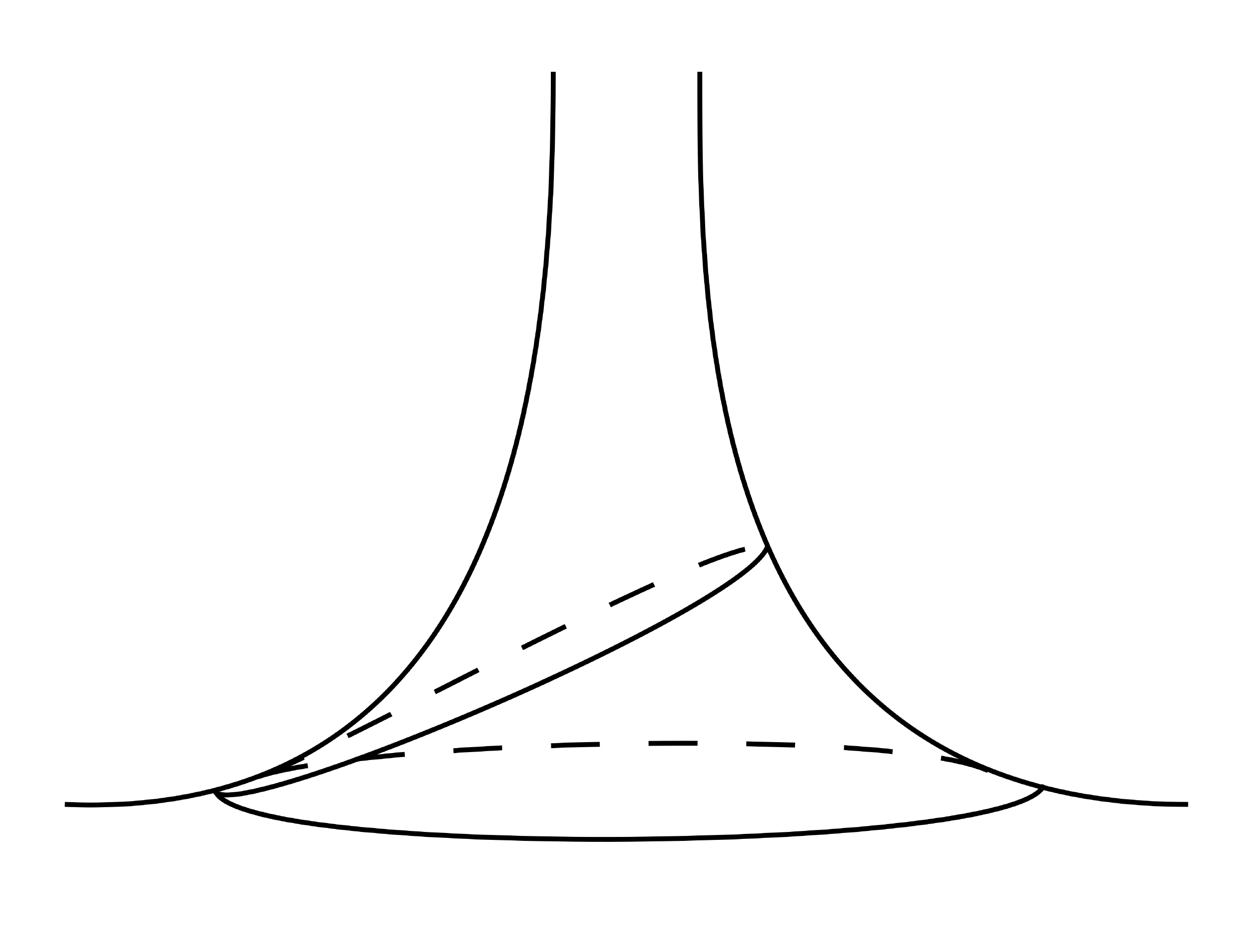}
\end{center}
\caption{The embedding of a throat in 3-dimensional Euclidean space qualitatively shows why a circle around the throat is not the shortest periodic path given a fixed starting point.}
\label{throat}
\end{figure}

One can understand this property pictorially by embedding a section of one of the throats in Euclidean 3-dimensional space (see Fig.~\ref{throat}). Note that the axionic direction is the periodic direction around the throat. The shape of the throat is essentially the reason why a simple closed circle around it does not provide the shortest path connecting a point to itself. Instead, we can minimize the length of this circle by pushing it upwards where the circumference of the throat with respect to the embedding space is smaller.

In summary, in spite of the possible $N$-fold widening of one or several throats by the flux, the field space increases only logarithmically with $N$.

\section{Size of the moduli space}

In the following we want to analyze our model from a four-dimensional point of view. The idea is to consider the four-dimensional effective field theory that describes the physics of our model at energy scales smaller than a cutoff $\Lambda$, and to determine the regions of moduli space where this theory is valid, i.e.~where KK- and winding modes are heavier than the cutoff scale. Once this region has been determined we will introduce a quantitative measure for the size of this region and formulate a conjecture about the dependence of this size on the cutoff.

\subsection{Winding and KK modes on the compact space}

Consider the $i$th of our three tori with complex structure modulus $\tau_i$. In (\ref{latticevec}) we have introduced the basis vectors $e_{i,x}=1$ and $e_{i,y}=\tau_i$ spanning the corresponding lattice in the complex plane. So far, no information concerning the volume is provided. By multiplying $e_{i,x}$ and $e_{i,y}$ by a factor $\sqrt{V_i/\text{Im}\,\tau_i}$ we obtain the basis vectors for a lattice corresponding to a torus with volume $V_i$:
\begin{equation}
\sqrt{\frac{V_i}{\text{Im}\,\tau_i}}\quad\text{and}\quad\sqrt{\frac{V_i}{\text{Im}\,\tau_i}}\tau_i\,.
\label{windlat}
\end{equation}
These vectors determine the mass $M_\text{W}$ of the winding modes on this torus via the formula\footnote{The prefactor $(2\pi\alpha')^{-1}$ comes from the Nambu-Goto action $S_\text{NG}=(2\pi\alpha')^{-1}\int_\text{WS}$.}
\begin{equation}
M_\text{W}(n_x,n_y)=\frac{1}{2\pi\alpha'}\sqrt{\frac{V_i}{\text{Im}\,\tau_i}}|n_xe_{i,x}+n_ye_{i,y}|=2\pi\sqrt{\frac{V_i}{\text{Im}\,\tau_i}}|n_x+n_y\tau_i|
\end{equation}
with integers $n_x$ and $n_y$. In the last step we switched to units defined by $l_\text{s}=2\pi\sqrt{\alpha'}=1$. Analogously, the dual lattice is spanned by the vectors
\begin{equation}
\frac{\text{i}}{\sqrt{V_i\,\text{Im}\,\tau_i}}\quad\text{and}\quad\frac{-\text{i}}{\sqrt{V_i\,\text{Im}\,\tau_i}}\tau_i\,,
\end{equation}
and determines the masses $M_\text{KK}$ of the KK-modes on the torus according to
\begin{equation}
M_\text{KK}(n_x,n_y)=2\pi\frac{1}{\sqrt{V_i\,\text{Im}\,\tau_i}}|n_x-n_y\tau_i|\,,
\end{equation}
with, again, integers $n_x$ and $n_y$. Substituting $n_y\rightarrow -n_y$ shows that the masses of KK- and winding modes differ only by a factor $V_i$. 

We achieve equality at the self-dual point $V_i=1$. This is a convenient choice as it simplifies the analysis regarding the effects of KK and winding modes on the cutoff of the theory. However, for $V_i=1$ certain 1-cycles in the geometry will necessarily become sub-stringy over large regions of the moduli space of $\tau_i$. In this case, unsuppressed instantons can arise if a string worldsheet or D-brane wraps cycles with sub-stringy volume. They may correct the 4d action, e.g.~the K\"ahler metric. Similarly, light 4d states (particles, strings etc.) can arise from string worldsheets or branes wrapped on small cycles. This may also lead to corrections or affect the value of the cutoff of the effective 4d theory. A complete analysis of the cutoff of the effective theory thus has to take into account KK modes, winding modes as well as instantons and other light states. For a simpler presentation, we will disentangle this as follows. First, in this section we will proceed with the study of the effects of KK and winding modes, working at the self-dual point $V_i=1$ for simplicity, but ignoring all other corrections and light states. Then we will remove any extra light states and unsuppressed instantons by increasing the volumes $V_i$ such that no sub-stringy cycles remain. As this will also affect the masses of KK and winding modes we will need to modify the analysis of this section, which we will explain in section \ref{sec:instsuppress}. It will turn out that this modification is technically straightforward. Having laid out our strategy, we now continue with the analysis for $V_i=1$.

Now we need to know the mass of the lightest winding mode on the $i$th torus, denoted by $m_{\text{W},i}$, which is equivalent to finding the shortest vector of the lattice spanned by the basis (\ref{windlat}).\footnote{In the following we will only talk about winding modes which in our setting have the same masses as the KK modes. In particular we have $m_{\text{W},i}=m_{\text{KK},i}$.} This problem is in general not solvable analytically and we will only provide an estimate. First of all, we can apply Minkowski's theorem to this two-dimensional lattice which will give an upper bound for the length of the shortest lattice vector. According to our choice $V_i=1$, the area of the parallelogram which is spanned by the basis (\ref{windlat}) is equal to unity.\footnote{Note that this volume is independent of the choice of basis.} Then the theorem states that any convex subset of $\mathbb{C}$ that is symmetric with respect to the origin and has a volume larger than four contains a non-zero lattice point. If we choose this subset to be a disk, we can conclude that the shortest lattice vector can not be longer than the radius of this disk. This implies an upper bound of order one for all three tori. 

However, there are regions in moduli space in which the true length of the shortest lattice vector is orders of magnitude smaller and we would vastly overestimate the part of moduli space where the low energy effective field theory is valid. We can improve this situation by analyzing two special regions in which we can find a much better estimate for the length of the shortest lattice vector.

Consider first $\text{Im}\,\tau_i\ge1$. We have to minimize $(n+m\text{Re}\,\tau_i)^2+(m\text{Im}\,\tau_i)^2 $ with $n,m\in\mathbbm{Z}$. For $m\ne 0$, this is larger than unity. For $m=0$ the minimum is clearly one, realized by the vector $(1,0)$. The corresponding physical length is $1/\sqrt{\text{Im}\,\tau_i}=m_{\text{W},i}/(2\pi)$.

Second, focus on $|\text{Re}\,\tau_i|\le\text{Im}\,\tau_i\ll1$. This always holds at the bottom of the central cusp of the fundamental domain of, for example, $\tau_1$ (see Fig.~\ref{fundomain12}). Once again we need to minimize $(n+m\text{Re}\,\tau_i)^2+(m\text{Im}\,\tau_i)^2 $. For $n\ne0$ the minimum is unity, obtained for $n=1$ and $m=0$. If $n=0$, the shortest lattice vector is simply $\tau_i$, the length of which is smaller than unity. The corresponding physical length is $|\tau_i|/\sqrt{\text{Im}\,\tau_i}\sim\sqrt{\text{Im}\,\tau_i}\ll1$, giving rise to $m_{\text{W},i}=2\pi\sqrt{\text{Im}\,\tau_i}$.

In fact, we can extend this result for $i=1$ to all the other cusps in the fundamental domain of $\tau_1$. Note that, in principle, we can distinguish the cusps due to the flux. However, right now we are only concerned with a pure lattice property, namely the shortest lattice vector, which does not depend on the rest of the physical situation. We can therefore safely ignore the fluxes. This allows us to use the original full modular invariance of the torus to shift all the cusps onto the central one. The result $m_{\text{W},1}=2\pi\sqrt{\text{Im}\,\tau_1}$ is hence not only valid in the central cusp but also in all the others.

Our complete result for the smallest winding mode mass therefore reads
\begin{equation}
m_{\text{W},i}\sim
\begin{cases}
2\pi/\sqrt{\text{Im}\,\tau_i}, & \text{for }\text{Im}\,\tau_i\ge1 \\
2\pi\sqrt{\text{Im}\,\tau_i}, & \text{for }\text{Re}\,\tau_i+n\le\text{Im}\,\tau_i\ll1 \\
2\pi		,		&			\text{else}
\end{cases}
\,,
\label{smallestmass}
\end{equation}
where the integer $n$ is chosen such that $\text{Re}\,\tau_i+n\in(-1/2,1/2]$.

\subsection{The restricted moduli space}
\label{sec:restricted}
Now we fix the cutoff scale $\Lambda$ with respect to which we want winding modes (and KK-modes) to be heavy, i.e.~$m_{\text{W},i}>\Lambda$ for all $i$. This condition is only satisfied on a subset of the moduli space which depends on $\Lambda$. We call this subset the restricted moduli space $\mathcal{M}(\Lambda)$ in the following. More precisely, since we take the four-dimensional point of view, we fix the ratio of the cutoff and the four-dimensional Planck scale $M_4$,
\begin{equation}
\epsilon\equiv\frac{\Lambda}{M_4}\,,
\end{equation}
where in our units $M_4=\sqrt{4\pi}g_\text{s}^{-1}$  and $g_\text{s}$ is the string coupling.\footnote{This is due to our choice $V_i=1$.} In the following we we will restrict ourselves to $g_\text{s}<1$ in order to stay in the perturbative regime. The monodromic moduli space is parametrized by $\{\tau_i\}$ with the vacuum condition imposed and restricted to the appropriate fundamental domains. The next step will now be to determine the region in moduli space that is compatible with the condition $m_{\text{W},i}>\Lambda$ for all $i$, i.e.~the restricted moduli space $\mathcal{M}(\Lambda)$. 

Let us start by considering $\tau_3$ which is just equal to the axio-dilaton $\tau$ according to the vacuum conditions (\ref{superpot}). The condition $g_\text{s}<1$ is then equivalent to $\text{Im}\,\tau_3=\text{Im}\,\tau>1$. According to (\ref{smallestmass}) we need to impose
\begin{equation}
\Lambda=\epsilon M_4=\epsilon\sqrt{4\pi}\text{Im}\,\tau<\frac{2\pi}{\sqrt{\text{Im}\,\tau}}\,,
\end{equation}
where we used $\tau=\tau_3$ and $\text{Im}\,\tau>1$. This gives a bound $\text{Im}\,\tau<(\pi^2/\epsilon)^{2/3}$ for the axio-dilaton. Taking into account the appropriate moduli space metric, this is of course consistent with the expected logarithmic growth of moduli space size with $1/\epsilon$. Indeed, we did not try to create long trajectories in the $\tau_3/\tau$-part of moduli space. To simplify our analysis, we will set $\text{Im}\,\tau=\text{Im}\,\tau_3=1$ from now on. In this way, we are certain that no light KK or winding modes arise from extreme values of $\tau$ and $\tau_3$.

Next consider $\tau_1$ and $\tau_2$. The vacuum condition for $M=1$ reads $\tau_1=N\tau_2$. We choose $\tau_1$ to parametrize the flat directions. Consider first the region defined by $\text{Im}\,\tau_1\ge 1$. The lightest mode on the first torus in this region has mass $2\pi/\sqrt{\text{Im}\,\tau_1}$ according to (\ref{smallestmass}). Requiring $\Lambda<2\pi/\sqrt{\text{Im}\,\tau_1}$ gives $\text{Im}\,\tau_1<(2\pi/\Lambda)^2$. The resulting bound on the fundamental domain in the complex $\tau_1$-plane can be visualized as a horizontal line coming down from infinity as we increase $\Lambda$ (see Figs.~\ref{conspace1} and \ref{conspace2}).

Now let us focus on the lower part of the moduli space, i.e.~$\text{Im}\,\tau_1<1$ and in particular on the cusps located near the real axis (see Fig.~\ref{fundomain12}). If we go far enough down the cusp we will always satisfy $|\text{Re}\,\tau_1|\le\text{Im}\,\tau_1$ (possibly after an integer shift along $\text{Re}\,\tau_1$) all the way to the singularity at the real axis. In fact, this condition covers most of the fundamental domain in the regime $\text{Im}\,\tau_1<1$ and we will therefore take the resulting constraint on the moduli space to be valid throughout this region. From (\ref{smallestmass}) we 
can read off the lightest winding mass coming from the first torus to be $2\pi\sqrt{\text{Im}\,\tau_1}$ which leads to the bound $\text{Im}\,\tau_1>(\Lambda/(2\pi))^2$. Similarly to the previously derived bound one can think of this as a horizontal line which now rises from the bottom of the cusps as we increase $\Lambda$. Our final picture of the restricted moduli space is sketched in Fig.~\ref{conspace1} and Fig.~\ref{conspace2}.

In the previous analysis we have glossed over a subtlety which we want to comment on in the following. So far we have ignored possible bounds coming from the second torus in the last two paragraphs. Now we argue that such bounds do not generically occur throughout the fundamental domain. Ignoring these additional but non-generic constraints will finally lead to an overestimation of the size of $\mathcal{M}(\Lambda)$. 

Let us concentrate on the region defined by $|\text{Re}\,\tau_1|\le\text{Im}\,\tau_1<N$. Then the vacuum condition $\tau_1=N\tau_2$ obviously implies $|\text{Re}\,\tau_2|\le\text{Im}\,\tau_2<1$.  According to (\ref{smallestmass}) we expect the lightest winding mass from the second torus to be $2\pi\sqrt{\text{Im}\,\tau_2}=2\pi\sqrt{\text{Im}\,\tau_1/N}$. In order to compare this with the corresponding winding masses of the first torus we need to differentiate between two cases. 

First focus on $\text{Im}\,\tau_2<1/N$, i.e.~$\text{Im}\,\tau_1<1$. The lightest winding mode on the first torus is then $2\pi\sqrt{\text{Im}\,\tau_1}$. This is heavier than the winding mode on the second torus which therefore provides the strongest bound on the moduli space. Second, consider $1/N\le\text{Im}\,\tau_2\le1$ or equivalently $1\le\text{Im}\,\tau_1\le N$. For $\text{Im}\,\tau_1<\sqrt{N}$ the lightest winding mode on the second torus is in fact lighter than the corresponding mode on the first torus, which has a mass $2\pi\sqrt{\text{Im}\,\tau_1}$. Consequently, the second torus would provide the most important bound on the moduli space in the regime $|\text{Re}\,\tau_1|\le\text{Im}\,\tau_1<\sqrt{N}$. 

However, the above region covers only the central cusp and a finite part of the upper region of the fundamental domain of $\tau_1$ which does not comprise a substantial part thereof.\footnote{One might be tempted to extend the validity of this bound to all cusps by using the shift symmetry of the second torus as was done in the last subsection for the first. The following argument shows why this is not possible: Consider a point in one of the cusps other than the central one. Then we have $|\text{Re}\,\tau_1|>\text{Im}\,\tau_1$ and hence also $|\text{Re}\,\tau_2|>\text{Im}\,\tau_2$. Now, in contrast to $\tau_1$, it is not possible to shift $\tau_2$ such that $|\text{Re}\,\tau_2|\le\text{Im}\,\tau_2$ holds because we already have $|\text{Re}\,\tau_2|\le1/2$ (remember that $|\text{Re}\,\tau_1|\le N/2$).} The corresponding additional bound can hence not be considered generic and may be safely omitted from our parametric analysis.

\subsection{Estimating the size of the restricted moduli space}
\label{sec:estimate}
Now we introduce a quantitative measure for the size of the restricted moduli space $\mathcal{M}(\Lambda)$. Since we are interested in distances in field space we may try to use the standard mathematical notion of the diameter. For a Riemannian manifold, in our case $\mathcal{M}(\Lambda)$, it is defined as
\begin{equation}
\text{diam}(\mathcal{M}(\Lambda))\equiv\sup_{p_1,p_2\in \mathcal{M}(\Lambda)}\inf_\gamma L_\gamma(p_1,p_2)\,,
\label{diam}
\end{equation}
where the infimum is taken over all curves $\gamma$ that connect the points $p_1$ and $p_2$ and $L_\gamma(p_1,p_2)$ denotes the length of the corresponding path. The quantity $d(p_1,p_2)\equiv\inf_\gamma L_\gamma(p_1,p_2)$ is the usual notion of distance between two points.\footnote{We will see below that in our physical situation this requires adjustment.} It is in particular extremal and the corresponding curve must hence be a geodesic. Note that an alternative measure for the size of $\mathcal{M}(\Lambda)$ is its volume which, however, we will not consider in the following.

\begin{figure}[t]
\begin{center}
\begin{tikzpicture}
\fill[black!10!white] (-5,0 ) rectangle (5,0.5);
\fill[black!10!white] (-5,5 ) rectangle (5,5.8);
\draw[-latex] (-5.5,0) -- (5.5,0) node[anchor = north west] {Re$\,\tau_1$};
\draw [-latex] (0,0) -- (0,6) node[anchor = south east] {Im$\,\tau_1$};
\draw (-5,0.1) -- (-5,-0.1) node[anchor=north] {$-\frac{N}{2}$};
\draw (5,0.1) -- (5,-0.1) node[anchor=north] {$\frac{N}{2}$};
\draw (0.1,1.5) -- (-0.1,1.5);
\draw (-0.2,1.5) node {1};
\draw (0.1,0.5) -- (-0.1,0.5) node[anchor=east] {$\left(\frac{\Lambda}{2\pi}\right)^2$};
\draw (0.1,5) -- (-0.1,5) node[anchor=east] {$\left(\frac{2\pi}{\Lambda}\right)^2$};
\draw [line width=0.4 mm] (-3,0.5) --(-3,5);
\draw [dashed,line width=0.4 mm] (0,0.5) -- (0,1.2);
\draw [dashed,line width=0.4 mm] (5,3) -- (5,5);
\draw [dashed,line width=0.4 mm] (5,3) arc (59.83:159.77:3.47);
\draw [line width=0.4 mm] (5,0.5) arc (11.3:168.7:2.55);
\filldraw (0,0.5) circle (0.05) node[anchor=west] {$B_1$};
\filldraw (5,0.5) circle (0.05) node[anchor=west] {$B_2$};
\filldraw (-3,5) circle (0.05) node[anchor=west] {$A_1$};
\filldraw (-3,0.5) circle (0.05) node[anchor=west] {$A_2$};
\filldraw (0,1.2) circle (0.05);
\filldraw (5,3) circle (0.05) node[anchor=west] {$C_2$};
\draw (-0.3,1) node {$C_1$};
\end{tikzpicture}
\end{center}
\caption{The constraints on the moduli space for $\Lambda<4\pi/\sqrt{N}$ are shown. This picture must be superposed with an appropriate fundamental domain of the congruence subgroup $\Gamma^0(N)$ in order to explicitly see the restricted moduli space. The grey shaded region is excluded by the lower and upper bounds given by $(\Lambda/(2\pi))^2$ and $(2\pi/\Lambda)^2$. In the text we calculate the length of the paths shown.}
\label{conspace1}
\end{figure}
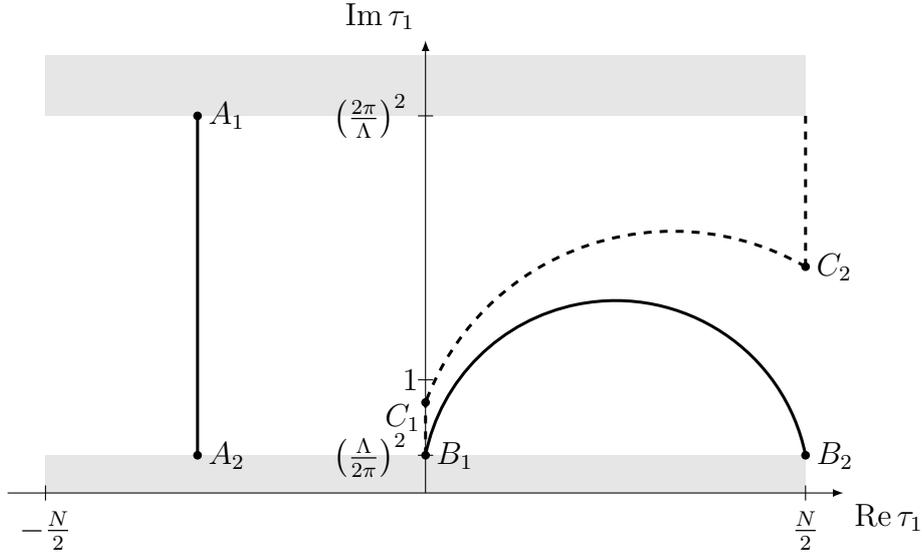

The technical task now is to estimate the diameter of $\mathcal{M}(\Lambda)$. For the unrestricted moduli space $\mathcal{M}(0)$ it is obvious that points, e.g.~in two different throats, can have an arbitrarily large distance, see Fig.~\ref{riemann}. This is due to the fact that the throats are infinitely long. Now consider $\mathcal{M}(\Lambda)$ with a small $\Lambda$. We will see in a moment that the technical condition is $\Lambda<4\pi/\sqrt{N}$. In this case, Fig.~\ref{conspace1} applies. Here we explicitly see that the bounds cut the infinitely long throats. The most widely separated points are still two points in different throats, now pushed up the throat as far as allowed by the bounds.

We have to take two cases into account. Remember that the point at infinity in the $\tau_1$-plane as well as the cusps at the bottom of the fundamental domain correspond to throats. Connecting a point $A_1$ in the upper throat to a point $A_2$ in one of the throats at the bottom yields a potentially long geodesic which is drawn in Fig.~\ref{conspace1} as a vertical line. The length of this geodesic is, according to (\ref{linelength}), given by 
\begin{equation}
d(A_1,A_2)=2\sqrt{2}\ln\left(\frac{2\pi}{\Lambda}\right)\,.
\label{L1}
\end{equation}
The second possibility is to consider two points $B_1$ and $B_2$ which lie in two different cusps, i.e.~in two throats at the bottom of the fundamental domain. They are connected by an arc-shaped geodesic as shown in Fig.~\ref{conspace1}. Using (\ref{symmarc}) the length $d(B_1,B_2)$ of this path can be estimated by 
\begin{equation}
d(B_1,B_2)=2\sqrt{2}\ln\left(\frac{2\pi}{\Lambda}\right)+\sqrt{2}\ln\left(\frac{N}{2}\right)\,,
\label{L2}
\end{equation}
which is clearly larger than $d(A_1,A_2)$. Hence we conclude that for $\Lambda<4\pi/\sqrt{N}$ the diameter of the moduli space is bounded by $2\sqrt{2}\ln(2\pi/\Lambda)+\sqrt{2}\ln(N/2)$. 

Note that, in principle, the distance between the two points lying in different cusps may actually be smaller than this. It is conceivable that, due to the complicated topology of the central part of $\mathcal{M}(\Lambda)$, a shortcut between the two throats exists which has a length much below $2\sqrt{2}\ln(2\pi/\Lambda)+\sqrt{2}\ln(N/2)$. However, taking (\ref{L1}) into account, the diameter of moduli space can not be smaller than $2\sqrt{2}\ln(2\pi/\Lambda)$.

\begin{figure}[t]
\begin{center}
\begin{tikzpicture}
\fill[black!10!white] (-5,0 ) rectangle (5,1);
\fill[black!10!white] (-5,2.25 ) rectangle (5,5.8);
\draw[-latex] (-5.5,0) -- (5.5,0) node[anchor = north west] {Re$\,\tau_1$};
\draw [-latex] (0,0) -- (0,6) node[anchor = south east] {Im$\,\tau_1$};
\draw (-5,0.1) -- (-5,-0.1) node[anchor=north] {$-\frac{N}{2}$};
\draw (5,0.1) -- (5,-0.1) node[anchor=north] {$\frac{N}{2}$};
\draw (0.1,1.5) -- (-0.1,1.5);
\draw (-0.2,1.5) node {1};
\draw (0.1,2.25) -- (-0.1,2.25) node[anchor=east] {$\left(\frac{2\pi}{\Lambda}\right)^2$};
%\draw [line width=0.4mm](5,1) -- (5,1.5) arc (31:50.5:2.92);
%\draw [line width=0.4mm](0,1) -- (0,1.5) arc (149:129.5:2.92);
\draw [line width=0.4 mm] (5,1) arc (21.8:56.5:2.69);
\draw [line width=0.4 mm] (0,1) arc (158.2:123.5:2.69);
\draw [line width=0.4 mm,dashed] (4,1) arc (26.4:153.6:2.25);
\draw[line width=0.4mm] (-3,1) -- (-3,2.25);
\draw (0.1,1) -- (-0.1,1) node[anchor=east] {$\left(\frac{\Lambda}{2\pi}\right)^2$};
\filldraw (-3,2.25) circle (0.05) node[anchor=west] {$A_1$};
\filldraw (-3,1) circle (0.05) node[anchor=west] {$A_2$};
\filldraw (0,1) circle (0.05) node[anchor=west] {$B_1$};
\filldraw (5,1) circle (0.05) node[anchor=west] {$B_2$};
%\filldraw (0,1.5) circle (0.05) node[anchor=west] {$C_1$};
%\filldraw (5,1.5) circle (0.05) node[anchor=west] {$C_2$};
\filldraw (4,1) circle (0.05) node[anchor=west] {$B_2'$};
%\draw [line width=0.4mm](1.68,0) ++(139:2.25) arc (139:41:2.25) -- ++(0,-0.5);
\end{tikzpicture}
\end{center}
\caption{The constraints on the moduli space for $\Lambda\ge4\pi/\sqrt{N}$ are shown. The grey shaded region is excluded by the lower and upper bounds given by $(\Lambda/(2\pi))^2$ and $(2\pi/\Lambda)^2$. The upper bound cuts off a part of the arc-shaped geodesic connecting $B_1$ with $B_2$.}
\label{conspace2}
\end{figure}
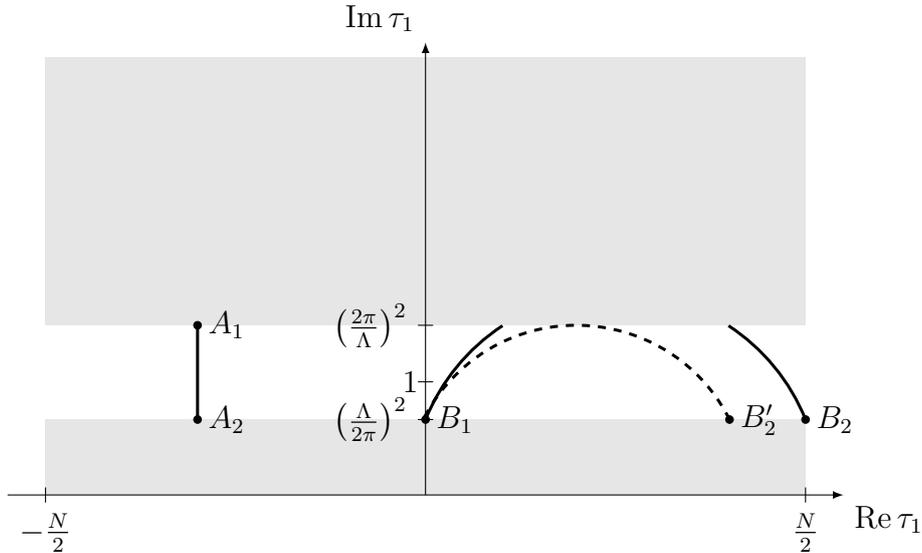

Next consider $\Lambda\ge4\pi/\sqrt{N}$. This situation is depicted in Fig.~(\ref{conspace2}). The formula for the distance between $A_1$ and $A_2$ remains the same as in the previous discussion. However, in the figure one can see that the upper bound cuts part of the arc-shaped geodesic between $B_1$ and $B_2$. It is therefore not a path that determines the distance between its two endpoints any more. Instead, according to our original definition of distance, we must deform it in such a way that it lies completely within $\mathcal{M}(\Lambda)$ and has minimal length. This procedure will, however, lead to an increased distance between the points $B_1$ and $B_2$ because any deformation of this geodesic will increase its length. From a physical point of view this behavior is contrary to our expectation that $\text{diam}(\mathcal{M}(\Lambda))$ is a monotonically decreasing function of $\Lambda$. In the following we present two different meaningful modification of our definition of distance that are free of this drawback.

Note first that the 4d field theory with cutoff $\Lambda$ breaks down at the boundary of $\mathcal{M}(\Lambda)$. Let us take the four-dimensional point of view and assume that, also outside this boundary, a meaningful 4d physical theory exists. In general, it ceases to be a local field theory and we are unable to make definite statements about the geometry of a corresponding larger moduli space. The most conservative approach is then to assume that all unknown distances are zero, in particular, that all pairs of boundary points have zero distance.

This idea can be made mathematically more rigorous. We know that $\mathcal{M}(\Lambda)$ is a subset of the full moduli space $\mathcal{M}(0)$. However, this may be only one of many manifolds of which $\mathcal{M}(\Lambda)$ could in principle be a subset. Let us denote by $\Omega(\Lambda)$ the set of all manifolds $\mathcal{M}$ such that $\mathcal{M}(\Lambda)\subset\mathcal{M}$ as a metric manifold. One can think of $\Omega(\Lambda)$ as parametrizing our ignorance about the true $\mathcal{M}(0)$ as a four-dimensional observer constrained by $\Lambda$. Our proposal for a new definition of a distance $d^*(p_1,p_2)$ between points $p_1,p_2\in\mathcal{M}(\Lambda)$ is
\begin{equation}
d^*(p_1,p_2)\equiv\inf_{\mathcal{M}\in\Omega(\Lambda)}d_\mathcal{M}(p_1,p_2)\,,
\end{equation}
where $d_\mathcal{M}$ is the usual distance on $\mathcal{M}$ and points in $\mathcal{M}(\Lambda)$ may be identified with points in $\mathcal{M}$ via an appropriate injection $i:\mathcal{M}(\Lambda)\rightarrow\mathcal{M}$. We expect that points at the boundary of $\mathcal{M}(\Lambda)$ are arbitrarily close in an appropriate $\mathcal{M}\in\Omega(\Lambda)$ which leads to the procedure of effectively compactifying all boundary points of $\mathcal{M}(\Lambda)$ to a single point, as was described in the previous paragraph. 

In the foregoing discussion we motivated the definition of $d^*$ by assuming that $\mathcal{M}(\Lambda)$ is part of a larger and more complete moduli space. Now we want to take the more radical point of view that, as 4d observers constrained by $\Lambda$, we are not allowed to venture outside the boundary even in principle. It may then be natural to work with a distance
\begin{equation}
d^\#(p_1,p_2)\equiv
\begin{cases}
d(p_1,p_2),	&	\text{if }p_1\text{ and }p_2\text{ are connected by a geodesic} \\
\text{undefined,}		&	\text{else}
\end{cases}\,,
\end{equation}
i.e.~to assume that points which are not connected by a geodesic that completely lies within $\mathcal{M}(\Lambda)$ do not have a well-defined distance and are treated as completely unrelated.
In a sense this definition of distance is much simpler and straightforward than our first proposal. The diameter of a general $\mathcal{M}(\Lambda)$, however, does not necessarily have to be a monotonically decreasing function of $\Lambda$ with this definition of distance, although this problem does not arise in our concrete example.

Now that we have discussed two different modified definitions of distance that are better suited to the problem at hand than the usual definition, we have to re-examine the analysis we have already worked out for $\Lambda<4\pi/\sqrt{N}$. The main difference between $d$ and $d^*$ is that all boundary points are identified to a single point if we use the latter. In particular, this implies that e.g.~a point at the upper and a point at the lower boundary in Fig.~\ref{conspace1} have zero distance. Therefore points at different boundaries are no longer good candidates for a large distance. 

Instead, potentially large distances can be achieved between points $C_1$ and $C_2$ (see Fig.~\ref{conspace1}). These are connected by the dashed arc-shaped geodesic as well as by the two dashed vertical geodesics and the boundary. Altogether these three different paths build a closed curve on which $C_1$ and $C_2$ lie. The maximal distance $d^*(C_1,C_2)$ is achieved if the length of the arc-shaped geodesic equals the sum of the lengths of the two vertical lines and at the same time is maximized. Since the analytic solution of this optimization problem is rather cumbersome, we give a qualitative discussion in three different parametric regimes: $2\pi/\Lambda \gg N/2$, $\,\,\, N/2 \gg 2\pi/\Lambda \gg \sqrt{N}/2\,$, and $\,\sqrt{N}/2\gg 2\pi/\Lambda$. We expect the result to capture the essential behavior of $\text{diam}^* (\mathcal{M} (\Lambda))$.\footnote{$\text{diam}^*$ and $\text{diam}^\#$ are defined as in (\ref{diam}) but using $d^*$ and $d^\#$, respectively, as the distance instead of $d$.}

Let us start in the regime $2\pi/\Lambda\gg N/2$. Then the contribution to the length of the arc-shaped geodesic due to its horizontal extension is completely negligible compared to the vertical direction (cf.~(\ref{L2})). Therefore, the length of the closed dashed path is to good accuracy given by $2\sqrt{2}\ln(2\pi/\Lambda)$ where we have only taken the vertical direction into account. At the optimum, $C_1$ and $C_2$ divide the path in two equally long parts such that their distance is
\begin{equation}
d^*(C_1,C_2)=\sqrt{2}\ln\left(\frac{2\pi}{\Lambda}\right)\quad\text{for}\quad\frac{2\pi}{\Lambda}\gg \frac{N}{2}\,.
\end{equation}

As we increase $\Lambda$ the contribution of the horizontal direction to the arc-shaped path becomes more and more important. According to (\ref{L2}), it can be estimated by $\sqrt{2}\ln(N/2)$, such that it starts to dominate at $2\pi/\Lambda\sim N/2$. Hence, in the regime $N/2 \gg 2\pi/\Lambda \gg \sqrt{N}/2$, the distance $d^*(C_1,C_2)$ is dominated by the length $\sim \sqrt{2}\ln(N/2)$ of the arc-shaped path. In this regime, the vertical positions of $C_1$ and $C_2$ keep adjusting as $\Lambda$ grows such that the vertical path maintains the same length.

The next qualitative change occurs when $\Lambda$ has increased so much that $\sqrt{N}/2\sim 2\pi/\Lambda$. Now the arc-shaped geodesic is cut by the upper bound and is hence no longer available in the competition with the vertical path. The vertical positions of $C_1$ and $C_2$ have by now moved to Im$\,\tau_1 = 1$, where they will stay from now on. Their distance is determined by the corresponding vertical geodesics connecting them to the lower and upper boundary respectively. Thus, in the new regime $\sqrt{N}/2\gg 2\pi/\Lambda$, this distance is $2\sqrt{2}\ln(2\pi/\Lambda)$. Combining the three regimes we have
\begin{equation}
\text{diam}^*(\mathcal{M}(\Lambda))\sim
\begin{cases}
\sqrt{2}\ln\left(\frac{2\pi}{\Lambda}\right)    &    \text{for }\Lambda\ll 4\pi/N    \\
\sqrt{2}\ln\left(\frac{N}{2}\right)    &    \text{for }4\pi/N \ll\Lambda\ll4\pi/\sqrt{N}\\
2\sqrt{2}\ln\left(\frac{2\pi}{\Lambda}\right)    &    \text{for }\Lambda\gg4\pi/\sqrt{N}
\end{cases}
\,.
\end{equation}

Finally we have to repeat this analysis for $d^\#$. For $\Lambda<4\pi/\sqrt{N}$ our original analysis remains valid and the diameter of $\mathcal{M}(\Lambda)$ is estimated by $\text{diam}^\#(\mathcal{M}(\Lambda))=2\sqrt{2}\ln(2\pi/\Lambda)+\sqrt{2}\ln(N/2)$. Once $\Lambda\ge4\pi/\sqrt{N}$ the arc-shaped geodesic is cut into two parts and the points $B_1$ and $B_2$ are no longer connected by a geodesic (see Fig.~\ref{conspace2}). Widely separated points that have a well-defined distance are now given by $B_1$ and $B_2'$ which are connected by the path shown in Fig.~\ref{conspace2}. Similarly to our original discussion this path provides an upper bound for the distance of the two points. In particular, the radius of the arc-shaped part is equal to $(2\pi/\Lambda)^2$. With (\ref{symmarc}) we calculate the length of this path to be $2\sqrt{2}\ln(2\pi/\Lambda)+2\sqrt{2}\ln(\sqrt{2}\pi/\Lambda)$. Hence, the diameter of $\mathcal{M}(\Lambda)$ reads
\begin{equation}
\label{eq:diamresult} \text{diam}^\#(\mathcal{M}(\Lambda))\sim
\begin{cases}
2\sqrt{2}\ln\left(\frac{2\pi}{\Lambda}\right)+\sqrt{2}\ln\left(\frac{N}{2}\right)	&	\text{for } \Lambda<\frac{4\pi}{\sqrt{N}}	\\
2\sqrt{2}\ln\left(\frac{2\pi}{\Lambda}\right)+2\sqrt{2}\ln\left(\frac{2\sqrt{2}\pi}{\Lambda}\right)	&	\text{for } \Lambda\ge\frac{4\pi}{\sqrt{N}}
\end{cases}
\,.
\end{equation}

Summarizing, we have found that the diameter of the restricted moduli space $\mathcal{M}(\Lambda)$  is estimated by $\ln(1/\Lambda)$ if we ignore order one pre-factors. Remarkably, this was found independently for two different definitions of distance. This is exactly the logarithmic behavior known from the Swampland conjecture. However, in our case we have a statement about the absolute size of the restricted moduli space instead of a statement about the relative size of KK and winding mode masses at two different points with a given distance. 

Before formulating our conjecture let us return to the problem of sub-stringy cycles. The analysis so far has been performed at the self-dual point with all torus volumes chosen to be $V_i=1$. As a result we cannot avoid cycles with sub-stringy volumes which in turn gives rise to unsuppressed contributions from both worldsheet and brane instantons. To arrive at a robust result for the diameter of $\mathcal{M}(\Lambda)$ these effects need to be accounted for. This is the subject of the next section.

\subsection{Suppression of worldsheet instantons}
\label{sec:instsuppress}
So far we have neglected the effect of worldsheet and brane instantons on our discussion of the size of moduli space. To ensure that we can safely ignore instanton effects, we need to arrange for the geometry not to possess any cycles with sub-stringy volumes. All cycles have to be super-stringy (which is equivalent to requiring that all winding masses are larger than $2\pi$). Most importantly, this can always be achieved by increasing the torus volumes $V_i$ sufficently. Here we analyse how this will affect the size of the moduli space.\footnote{We thank the referee for prompting us to add this discussion.}

Let us first consider the third torus. Recall that we have set $\text{Im}\,\tau_3=1$ such that $V_3=1$ suffices according to (\ref{windlat}) to make both cycles of $T_3^2$ have exactly string length. However, for the first torus we have to increase the volume $V_1$ to ensure that both cycles on $T_1^2$ are super-stringy. In particular, we require
\begin{equation}
V_1=
\begin{cases}
\text{Im}\,\tau_1, & \text{for }\text{Im}\,\tau_1\ge1 \, ,\\
1/\text{Im}\,\tau_1, & \text{for }\text{Im}\,\tau_1<1 \, .
\end{cases}
\label{volume}
\end{equation}
Now let us turn to torus $T_2^2$. At the end of Section 4.2 we have argued that the winding masses coming from the second torus are generically larger than the ones from $T_1^2$. The argument was made for $V_{1,2}=1$ but remains true for the more general situation $V_1=V_2$, as is evident from (\ref{smallestmass}). Therefore, by choosing $V_2=V_1$ with $V_1$ given by (\ref{volume}), we find that $m_{\text{W},2}>2\pi$. This ensures that both cycles on $T_2^2$ are super-stringy, at least generically. With these choices for the volumes $V_i$ no sub-stringy cycles remain and instantons can be safely ignored.

There are two points in the analysis in Sections 4.1 to 4.3 that need to be modified because of our different choice of volumes. First of all, as we have increased the winding masses beyond the self dual point we also decreased the masses of KK modes accordingly. Hence the KK modes now give rise to the stronger constraints on the validity of the 4d effective theory. Inserting a factor $1/\sqrt{V_1}$ in (\ref{smallestmass}) with $V_1$ as in (\ref{volume}) we find for the smallest KK mass
\begin{equation}
m_{\text{KK},1}\sim
\begin{cases}
2\pi/\text{Im}\,\tau_1, & \text{for }\text{Im}\,\tau_1\ge1 \\
2\pi\text{Im}\,\tau_1, & \text{for }\text{Re}\,\tau_1+n\le\text{Im}\,\tau_1\ll1 \\
2\pi		,		&			\text{else}
\end{cases}
\,.
\end{equation}
Demanding $m_{\text{KK},1}>\Lambda$ we find that the horizontal lines in Figs. \ref{conspace1} and \ref{conspace2} are no longer at $(\Lambda/(2\pi))^2$ and $(2\pi/\Lambda)^2$ but at $\Lambda/(2\pi)$ and $2\pi/\Lambda$, respectively. Consequently, all expressions regarding the size of the moduli space have to be modified by substituting $(\Lambda/(2\pi))^2\rightarrow\Lambda/(2\pi)$. Note that this replacement does not change the formulae for the diameter significantly since the cutoff $\Lambda$ always appears within a logarithm.

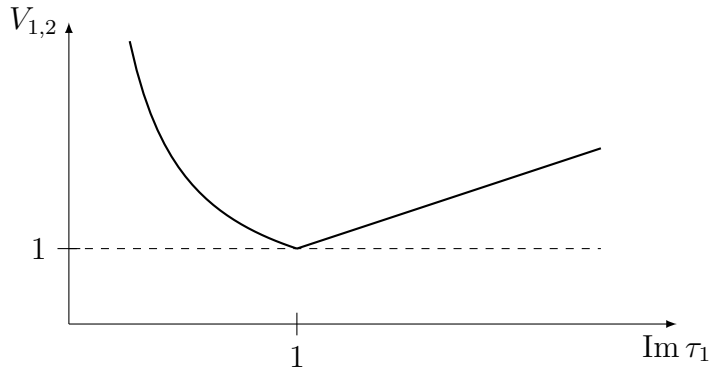
\begin{figure}[t]
\begin{center}
\begin{tikzpicture}[domain=0.8:3]

\draw[-latex] (0,0) --(8,0) node[anchor=north] {$\text{Im}\,\tau_1$};
\draw[-latex] (0,0) --(0,4) node[anchor=east] {$V_{1,2}$};
\draw (3,0.15) -- (3,-0.15) node[anchor=north] {1};
\draw (0.15,1) -- (-0.15,1) node[anchor=east] {1};
\draw[thick] plot (\x,{3/\x});
\draw[thick] (3,1) -- (7,2.33);
\draw [dashed] (0,1) -- (7,1);

\end{tikzpicture}
\end{center}
\caption{The dashed line corresponds to the submanifold of moduli space that is parametrized by $\tau_1$ for constant torus volumes $V_i$. Letting the volumes $V_i$ depend on $\tau_1$ as in (\ref{volume}) gives rise to a different submanifold denoted by the solid line. Distances within this submanifold can to be calculated by considering the contributions from both the metric on complex structure and K\"ahler moduli space.}
\label{volumesketch}
\end{figure}

Furthermore we should take into account that $V_{1,2}$ are no longer constant and therefore contribute to the distance traversed in moduli space as we vary $\tau_1$. Indeed, we have so far discussed distances in a submanifold of moduli space defined by fixing the $V_i$ and $\tau_3$ and only varying $\tau_1=N\tau_2$. By contrast, we now have to consider a submanifold which is non-trivially embedded in the product of K\"ahler and complex structure moduli spaces as sketched in Fig.~\ref{volumesketch}. The contribution to the metric on this submanifold due to the displacement of K\"ahler moduli can be calculated from the corresponding K\"ahler potential. For the K\"ahler moduli sector this is given by
\begin{equation}
\mathcal{K}=-\ln\left[\frac{1}{8}(T_1+\bar{T}_1)(T_2+\bar{T}_2)(T_3+\bar{T}_3)\right]+\ldots \, , \quad \textrm{with} \quad \textrm{Re}(T_1)=V_2V_3 \, , \quad \textrm{etc}. 
\end{equation}
Using this and (\ref{volume}) we find for the metric of the subset of the full moduli space parametrized by $\tau_1$:
\begin{equation}
ds^2= 2 \, \frac{\text{d}\tau_1\text{d}\overline{\tau_1}}{(\text{Im}\,\tau_1)^2} \, .
\end{equation}
In our original and simplified analysis the metric (cf.~(\ref{modmetric1})) was smaller by a factor four with the corresponding distances smaller by a factor of two. Recall that the replacement $(\Lambda/(2\pi))^2\rightarrow\Lambda/(2\pi)$ introduced a factor $1/2$ in those terms in \eqref{eq:diamresult} which involve a logarithm of $\Lambda$. This factor is cancelled by the additional factor two from the new metric, such that the sole net effect is the substitution $\ln N\to 2\ln N$ plus non-logarithmic terms.  Thus, the introduction of variable volumes does not change our final formulae \eqref{eq:diamresult} for the diameter of the moduli space significantly.

\subsection{Statement of the conjecture}

\emph{Consider a 4d field theory with cutoff $\Lambda$. The diameter of the corresponding moduli space (as defined in section \ref{sec:estimate}) is then of the order $\sim\ln(1/\Lambda)$.}

This formulation is very natural if one is interested in long flat directions in moduli space in the absence of potentials. For example, if one is interested in effective field theories for large-field inflation, this theory must be valid at least at the energy scale of inflation given by $H$. Our conjecture implies then that flat directions have at most lengths of the order $\ln(1/H)$. Note that this statement is true only in the absence of potentials and it therefore does not automatically rule out models of large-field inflation with too large $H$.

Another conjecture which is closer to the original Swampland conjecture is:

\emph{Consider the moduli space of a string theory compactification to four dimensions. Consider two points in this space with a distance $L$ determined by a certain geodesic. Then there exist points on this geodesic at which the lightest KK or winding mode mass is below or of the order $\exp(-\alpha L)$, with $\alpha\sim{\cal O}(1)$.}

A subtle but practically important difference to the Swampland Conjecture is the following: According to our conjecture it is possible to have two points in moduli space which have a large distance and, at the same time, KK and winding modes of the same, high masses. The low-mass or low cutoff situation occurs somewhere in between. This is in particular what happens for points separated in the axionic coordinate (i.e.~$\textrm{Re} \, \tau$ in our explicit model). The lowest cutoff will be experienced at a point along the geodesic connecting the two points, and not at either the beginning or endpoint.

\section{Conclusions}
In this work we examined the possibility of trans-Planckian field spaces for complex structure moduli in string compactifications employing toroidal orientifolds. The main observation is that by a suitable choice of 3-form fluxes, a certain combination of moduli is lifted, such that the remaining complex flat direction exhibits an enlarged fundamental domain compared to the canonical fundamental domain of a complex structure modulus of a torus. We refer to this as a `Monodromic Moduli Space'. 

Mathematically, this moduli space corresponds to the fundamental domain of a congruence subgroup of SL$(2, \mathbbm{Z})$. One important observation is that the fundamental domain of such a congruence subgroup is typically widened compared to the canonical fundamental domain of SL$(2, \mathbbm{Z})$. This widening takes the form
\begin{align}
\textrm{Re} \, \tau \in [- \tfrac{1}{2}, \tfrac{1}{2}] \quad \longrightarrow \quad \textrm{Re} \, \tau \in [- \tfrac{N}{2}, \tfrac{N}{2}] \, ,
\end{align}
where $\tau$ is a complex structure modulus and $N$ is an integer set by flux numbers. 

We proceeded by examining whether a Monodromic Moduli Space may allow 
for trans-Planckian field displacements. First we note that `axionic' 
trajectories, i.e.~trajectories with Im$\,\tau =\,$const., can become 
large to the extent that $N$ can. (The tadpole constraint on 3-form 
fluxes implies $N \leq 16$ in our toy model.) But second we also note 
that for any two points on such a long (non-geodesic) trajectory much 
shorter connections exist. They correspond to arcs in the hyperbolic 
plane and their length scales only as $\ln N$. Moreover, we can restrict 
our moduli space by demanding that no winding or KK modes appear below a 
certain cutoff $\Lambda$. It then turns out that an appropriately 
defined maximal distance between points on an axionic trajectory is not 
only bounded by $\ln N$ but also by $\ln(1/\Lambda)$. This is 
reminiscent of the logarithmic limitations of field ranges due to 
backreaction observed in \cite{1602.06517}, but here a related 
phenomenon arises for flat directions.

While we made our observations in a simple string compactification based on a toroidal orientifold, we expect them to hold more widely. To be specific, monodromies also exist in flux compactifications on Calabi-Yau (CY) manifolds, an observation that has been exploited to study moduli dynamics and tunneling between different vacua in the string landscape, see e.g.~\cite{0612222, 0805.3705, 1011.6588, 1108.1394}. In our context, the key point is that CY moduli spaces have large complex structure points, analogous to the point at imaginary infinity in the torus fundamental domain. The simplest example is $(T^2)^3$, where we are dealing with the direct product of three of the familiar throat-like geometries. In general, the geometry near the large complex structure point of a CY is much more complicated, but it always includes `axionic' directions which characterize short paths around such points. These paths are periodic if one allows for identifications using large diffeomorphisms of the CY. We expect that this periodicity can be enlarged by an appropriate flux choice, analogously to our torus orientifold example. We also expect that the resulting long axionic trajectories will be very far from geodesics, with shortcuts similar to our arcs in the hyperbolic plane. Thus, the qualitative structure of a Monodromic Moduli Space of a CY with 3-form flux should be similar to what we found in this paper. In the context of inflation, discussions of the moduli space at large complex structure appeared e.g.~in \cite{1404.3711, 1405.0283, Blumenhagen:2014nba, Garcia-Etxebarria:2014wla, Bizet:2016paj}; for recent progress concerning global CY moduli spaces see \cite{Donagi:2017mhd}; for recent work on moduli spaces of CY 4-folds see \cite{1604.05325}. 

The above motivates two conjectures which are related, but distinct from the various forms of the Swampland Conjecture \cite{0509212, 0605264, 1610.00010, 1705.04328}. Given the moduli space of a generic 4d field theory with cutoff $\Lambda$, we conjecture that the absolute size of the moduli space, as measured by the appropriately defined diameter, scales as $\ln(1/\Lambda)$. Alternatively, we may focus on the full moduli space of a certain string compactification. Pick two points in this moduli space which are connected by a minimal geodesic with length $L$. Then we claim that there exist points on this geodesic at which the lightest KK or winding mode mass is smaller or of the order of $\exp(-\alpha L)$, with $\alpha\sim{\cal O}(1)$.

One of the key findings of our work is that our construction allows for trans-Planckian `axionic' directions which, however, are not geodesics. In particular, a trajectory along $\textrm{Re} \, \tau$ for fixed $\textrm{Im} \, \tau=1$ is a periodic direction with period $N / \sqrt{2}$. This can be moderately trans-Planckian despite the tadpole constraint on $N$. The upshot is that if it were possible to stabilize $\textrm{Im} \, \tau$ without completely destroying the structure of the Monodromic Moduli Space, our construction may constitute the first step towards a theory of a trans-Planckian axion.

This is relevant for cosmology where one open question is the compatibility of large field inflation and theories of quantum gravity. It has been suggested that large-field inflation can in principle be embedded in the complex structure moduli sector of string theory compactifications \cite{1404.3711, Abe:2014xja,1503.07912,Blumenhagen:2014nba, Garcia-Etxebarria:2014wla, Bizet:2016paj}, as long as there exists a trans-Planckian axionic direction. We suggest that Monodromic Moduli Spaces may be a promising starting point for the construction of such models. 

However, there are also obstacles to be overcome: To stabilize $\textrm{Im} \, \tau$, we require contributions to the potential which may interfere with the proposed simple structure of the Monodromic Moduli Space. Both for this stabilization and to construct a more realistic model of cosmology and particle physics, it is necessary to move beyond simple toroidal orientifolds. While, as noted above, we expect the general structure of the corresponding Monodromic Moduli Spaces of CYs to be similar, the details are far from clear. For example, symmetry structures replacing the modular group and instanton-type (in the mirror dual language) corrections which lift `axionic' directions non-perturbatively have to be studied. We leave these interesting questions for future work.

\subsection*{Acknowledgments}

We thank Thomas Grimm, Hans Jockers, Miguel Montero, Eran Palti, Eric Sharpe, Pablo Soler and Irene Valenzuela for helpful discussions. This work was supported by the DFG Transregional Collaborative Research Centre TRR 33 ``The Dark Universe'' and the HGSFP. LW acknowledges support from the Advanced ERC grant ``SM-grav'', No 669288.

\bibliography{moduliSpaceSize.bib}

\providecommand{\href}[2]{#2}\begingroup\raggedright\begin{thebibliography}{10}

\bibitem{0509212}
C.~Vafa, \emph{{The String landscape and the swampland}},
  \href{https://arxiv.org/abs/hep-th/0509212}{{\tt hep-th/0509212}}.

\bibitem{0601001}
N.~Arkani-Hamed, L.~Motl, A.~Nicolis and C.~Vafa, \emph{{The String landscape,
  black holes and gravity as the weakest force}},
  \href{http://dx.doi.org/10.1088/1126-6708/2007/06/060}{\emph{JHEP} {\bf 06}
  (2007) 060}, [\href{https://arxiv.org/abs/hep-th/0601001}{{\tt
  hep-th/0601001}}].

\bibitem{0605264}
H.~Ooguri and C.~Vafa, \emph{{On the Geometry of the String Landscape and the
  Swampland}},
  \href{http://dx.doi.org/10.1016/j.nuclphysb.2006.10.033}{\emph{Nucl. Phys.}
  {\bf B766} (2007) 21--33}, [\href{https://arxiv.org/abs/hep-th/0605264}{{\tt
  hep-th/0605264}}].

\bibitem{Freese:1990rb}
K.~Freese, J.~A. Frieman and A.~V. Olinto, \emph{{Natural inflation with pseudo
  - Nambu-Goldstone bosons}},
  \href{http://dx.doi.org/10.1103/PhysRevLett.65.3233}{\emph{Phys. Rev. Lett.}
  {\bf 65} (1990) 3233--3236}.

\bibitem{Silverstein:2008sg}
E.~Silverstein and A.~Westphal, \emph{{Monodromy in the CMB: Gravity Waves and
  String Inflation}},
  \href{http://dx.doi.org/10.1103/PhysRevD.78.106003}{\emph{Phys. Rev.} {\bf
  D78} (2008) 106003}, [\href{https://arxiv.org/abs/0803.3085}{{\tt
  0803.3085}}].

\bibitem{0808.0706}
L.~McAllister, E.~Silverstein and A.~Westphal, \emph{{Gravity Waves and Linear
  Inflation from Axion Monodromy}},
  \href{http://dx.doi.org/10.1103/PhysRevD.82.046003}{\emph{Phys. Rev.} {\bf
  D82} (2010) 046003}, [\href{https://arxiv.org/abs/0808.0706}{{\tt
  0808.0706}}].

\bibitem{1504.07551}
P.~W. Graham, D.~E. Kaplan and S.~Rajendran, \emph{{Cosmological Relaxation of
  the Electroweak Scale}},
  \href{http://dx.doi.org/10.1103/PhysRevLett.115.221801}{\emph{Phys. Rev.
  Lett.} {\bf 115} (2015) 221801},
  [\href{https://arxiv.org/abs/1504.07551}{{\tt 1504.07551}}].

\bibitem{1409.5793}
T.~Rudelius, \emph{{On the Possibility of Large Axion Moduli Spaces}},
  \href{http://dx.doi.org/10.1088/1475-7516/2015/04/049}{\emph{JCAP} {\bf 1504}
  (2015) 049}, [\href{https://arxiv.org/abs/1409.5793}{{\tt 1409.5793}}].

\bibitem{1412.3457}
A.~de~la Fuente, P.~Saraswat and R.~Sundrum, \emph{{Natural Inflation and
  Quantum Gravity}},
  \href{http://dx.doi.org/10.1103/PhysRevLett.114.151303}{\emph{Phys. Rev.
  Lett.} {\bf 114} (2015) 151303}, [\href{https://arxiv.org/abs/1412.3457}{{\tt
  1412.3457}}].

\bibitem{1503.00795}
T.~Rudelius, \emph{{Constraints on Axion Inflation from the Weak Gravity
  Conjecture}}, \href{http://dx.doi.org/10.1088/1475-7516/2015/09/020,
  10.1088/1475-7516/2015/9/020}{\emph{JCAP} {\bf 1509} (2015) 020},
  [\href{https://arxiv.org/abs/1503.00795}{{\tt 1503.00795}}].

\bibitem{Brown:2015iha}
J.~Brown, W.~Cottrell, G.~Shiu and P.~Soler, \emph{{Fencing in the Swampland:
  Quantum Gravity Constraints on Large Field Inflation}},
  \href{http://dx.doi.org/10.1007/JHEP10(2015)023}{\emph{JHEP} {\bf 10} (2015)
  023}, [\href{https://arxiv.org/abs/1503.04783}{{\tt 1503.04783}}].

\bibitem{Bachlechner:2015qja}
T.~C. Bachlechner, C.~Long and L.~McAllister, \emph{{Planckian Axions and the
  Weak Gravity Conjecture}},
  \href{http://dx.doi.org/10.1007/JHEP01(2016)091}{\emph{JHEP} {\bf 01} (2016)
  091}, [\href{https://arxiv.org/abs/1503.07853}{{\tt 1503.07853}}].

\bibitem{1506.03447}
B.~Heidenreich, M.~Reece and T.~Rudelius, \emph{{Weak Gravity Strongly
  Constrains Large-Field Axion Inflation}},
  \href{http://dx.doi.org/10.1007/JHEP12(2015)108}{\emph{JHEP} {\bf 12} (2015)
  108}, [\href{https://arxiv.org/abs/1506.03447}{{\tt 1506.03447}}].

\bibitem{Heidenreich:2015nta}
B.~Heidenreich, M.~Reece and T.~Rudelius, \emph{{Sharpening the Weak Gravity
  Conjecture with Dimensional Reduction}},
  \href{http://dx.doi.org/10.1007/JHEP02(2016)140}{\emph{JHEP} {\bf 02} (2016)
  140}, [\href{https://arxiv.org/abs/1509.06374}{{\tt 1509.06374}}].

\bibitem{1701.05572}
M.~J. Dolan, P.~Draper, J.~Kozaczuk and H.~Patel, \emph{{Transplanckian
  Censorship and Global Cosmic Strings}},
  \href{http://dx.doi.org/10.1007/JHEP04(2017)133}{\emph{JHEP} {\bf 04} (2017)
  133}, [\href{https://arxiv.org/abs/1701.05572}{{\tt 1701.05572}}].

\bibitem{1701.06553}
A.~Hebecker, P.~Henkenjohann and L.~T. Witkowski, \emph{{What is the Magnetic
  Weak Gravity Conjecture for Axions?}},
  \href{http://dx.doi.org/10.1002/prop.201700011}{\emph{Fortsch. Phys.} {\bf
  65} (2017) 1700011}, [\href{https://arxiv.org/abs/1701.06553}{{\tt
  1701.06553}}].

\bibitem{1610.00010}
D.~Klaewer and E.~Palti, \emph{{Super-Planckian Spatial Field Variations and
  Quantum Gravity}},
  \href{http://dx.doi.org/10.1007/JHEP01(2017)088}{\emph{JHEP} {\bf 01} (2017)
  088}, [\href{https://arxiv.org/abs/1610.00010}{{\tt 1610.00010}}].

\bibitem{1602.06517}
F.~Baume and E.~Palti, \emph{{Backreacted Axion Field Ranges in String
  Theory}}, \href{http://dx.doi.org/10.1007/JHEP08(2016)043}{\emph{JHEP} {\bf
  08} (2016) 043}, [\href{https://arxiv.org/abs/1602.06517}{{\tt 1602.06517}}].

\bibitem{1703.05776}
R.~Blumenhagen, I.~Valenzuela and F.~Wolf, \emph{{The Swampland Conjecture and
  F-term Axion Monodromy Inflation}},
  \href{https://arxiv.org/abs/1703.05776}{{\tt 1703.05776}}.

\bibitem{1705.04328}
E.~Palti, \emph{{The Weak Gravity Conjecture and Scalar Fields}},
  \href{https://arxiv.org/abs/1705.04328}{{\tt 1705.04328}}.

\bibitem{1404.3040}
F.~Marchesano, G.~Shiu and A.~M. Uranga, \emph{{F-term Axion Monodromy
  Inflation}}, \href{http://dx.doi.org/10.1007/JHEP09(2014)184}{\emph{JHEP}
  {\bf 09} (2014) 184}, [\href{https://arxiv.org/abs/1404.3040}{{\tt
  1404.3040}}].

\bibitem{1404.3542}
R.~Blumenhagen and E.~Plauschinn, \emph{{Towards Universal Axion Inflation and
  Reheating in String Theory}},
  \href{http://dx.doi.org/10.1016/j.physletb.2014.08.007}{\emph{Phys. Lett.}
  {\bf B736} (2014) 482--487}, [\href{https://arxiv.org/abs/1404.3542}{{\tt
  1404.3542}}].

\bibitem{1404.3711}
A.~Hebecker, S.~C. Kraus and L.~T. Witkowski, \emph{{D7-Brane Chaotic
  Inflation}},
  \href{http://dx.doi.org/10.1016/j.physletb.2014.08.028}{\emph{Phys. Lett.}
  {\bf B737} (2014) 16--22}, [\href{https://arxiv.org/abs/1404.3711}{{\tt
  1404.3711}}].

\bibitem{0303252}
T.~Banks, M.~Dine, P.~J. Fox and E.~Gorbatov, \emph{{On the possibility of
  large axion decay constants}},
  \href{http://dx.doi.org/10.1088/1475-7516/2003/06/001}{\emph{JCAP} {\bf 0306}
  (2003) 001}, [\href{https://arxiv.org/abs/hep-th/0303252}{{\tt
  hep-th/0303252}}].

\bibitem{0605206}
P.~Svrcek and E.~Witten, \emph{{Axions In String Theory}},
  \href{http://dx.doi.org/10.1088/1126-6708/2006/06/051}{\emph{JHEP} {\bf 06}
  (2006) 051}, [\href{https://arxiv.org/abs/hep-th/0605206}{{\tt
  hep-th/0605206}}].

\bibitem{1412.1093}
T.~C. Bachlechner, C.~Long and L.~McAllister, \emph{{Planckian Axions in String
  Theory}}, \href{http://dx.doi.org/10.1007/JHEP12(2015)042}{\emph{JHEP} {\bf
  12} (2015) 042}, [\href{https://arxiv.org/abs/1412.1093}{{\tt 1412.1093}}].

\bibitem{1203.5476}
J.~P. Conlon, \emph{{Quantum Gravity Constraints on Inflation}},
  \href{http://dx.doi.org/10.1088/1475-7516/2012/09/019}{\emph{JCAP} {\bf 1209}
  (2012) 019}, [\href{https://arxiv.org/abs/1203.5476}{{\tt 1203.5476}}].

\bibitem{Kaloper:2015jcz}
N.~Kaloper, M.~Kleban, A.~Lawrence and M.~S. Sloth, \emph{{Large Field
  Inflation and Gravitational Entropy}},
  \href{http://dx.doi.org/10.1103/PhysRevD.93.043510}{\emph{Phys. Rev.} {\bf
  D93} (2016) 043510}, [\href{https://arxiv.org/abs/1511.05119}{{\tt
  1511.05119}}].

\bibitem{1503.03886}
M.~Montero, A.~M. Uranga and I.~Valenzuela, \emph{{Transplanckian axions!?}},
  \href{http://dx.doi.org/10.1007/JHEP08(2015)032}{\emph{JHEP} {\bf 08} (2015)
  032}, [\href{https://arxiv.org/abs/1503.03886}{{\tt 1503.03886}}].

\bibitem{1607.06814}
A.~Hebecker, P.~Mangat, S.~Theisen and L.~T. Witkowski, \emph{{Can
  Gravitational Instantons Really Constrain Axion Inflation?}},
  \href{http://dx.doi.org/10.1007/JHEP02(2017)097}{\emph{JHEP} {\bf 02} (2017)
  097}, [\href{https://arxiv.org/abs/1607.06814}{{\tt 1607.06814}}].

\bibitem{Alonso:2017avz}
R.~Alonso and A.~Urbano, \emph{{Wormholes and masses for Goldstone bosons}},
  \href{https://arxiv.org/abs/1706.07415}{{\tt 1706.07415}}.

\bibitem{Cottrell:2016bty}
W.~Cottrell, G.~Shiu and P.~Soler, \emph{{Weak Gravity Conjecture and Extremal
  Black Holes}},  \href{https://arxiv.org/abs/1611.06270}{{\tt 1611.06270}}.

\bibitem{Hebecker:2017uix}
A.~Hebecker and P.~Soler, \emph{{The Weak Gravity Conjecture and the Axionic
  Black Hole Paradox}},  \href{https://arxiv.org/abs/1702.06130}{{\tt
  1702.06130}}.

\bibitem{1705.06287}
S.~Hod, \emph{{A proof of the weak gravity conjecture}}, {\emph{Int. J. Mod.
  Phys.} {\bf D26} (2017) 1742004},
  [\href{https://arxiv.org/abs/1705.06287}{{\tt 1705.06287}}].

\bibitem{1706.08257}
Z.~Fisher and C.~J. Mogni, \emph{{A Semiclassical, Entropic Proof of a Weak
  Gravity Conjecture}},  \href{https://arxiv.org/abs/1706.08257}{{\tt
  1706.08257}}.

\bibitem{Kim:2004rp}
J.~E. Kim, H.~P. Nilles and M.~Peloso, \emph{{Completing natural inflation}},
  \href{http://dx.doi.org/10.1088/1475-7516/2005/01/005}{\emph{JCAP} {\bf 0501}
  (2005) 005}, [\href{https://arxiv.org/abs/hep-ph/0409138}{{\tt
  hep-ph/0409138}}].

\bibitem{Dvali:2005an}
G.~Dvali, \emph{{Three-form gauging of axion symmetries and gravity}},
  \href{https://arxiv.org/abs/hep-th/0507215}{{\tt hep-th/0507215}}.

\bibitem{1503.07912}
A.~Hebecker, P.~Mangat, F.~Rompineve and L.~T. Witkowski, \emph{{Winding out of
  the Swamp: Evading the Weak Gravity Conjecture with F-term Winding
  Inflation?}},
  \href{http://dx.doi.org/10.1016/j.physletb.2015.07.026}{\emph{Phys. Lett.}
  {\bf B748} (2015) 455--462}, [\href{https://arxiv.org/abs/1503.07912}{{\tt
  1503.07912}}].

\bibitem{Saraswat:2016eaz}
P.~Saraswat, \emph{{Weak gravity conjecture and effective field theory}},
  \href{http://dx.doi.org/10.1103/PhysRevD.95.025013}{\emph{Phys. Rev.} {\bf
  D95} (2017) 025013}, [\href{https://arxiv.org/abs/1608.06951}{{\tt
  1608.06951}}].

\bibitem{1601.00647}
J.~P. Conlon and S.~Krippendorf, \emph{{Axion decay constants away from the
  lamppost}}, \href{http://dx.doi.org/10.1007/JHEP04(2016)085}{\emph{JHEP} {\bf
  04} (2016) 085}, [\href{https://arxiv.org/abs/1601.00647}{{\tt 1601.00647}}].

\bibitem{0612222}
U.~H. Danielsson, N.~Johansson and M.~Larfors, \emph{{The World next door:
  Results in landscape topography}},
  \href{http://dx.doi.org/10.1088/1126-6708/2007/03/080}{\emph{JHEP} {\bf 03}
  (2007) 080}, [\href{https://arxiv.org/abs/hep-th/0612222}{{\tt
  hep-th/0612222}}].

\bibitem{0805.3705}
M.~C. Johnson and M.~Larfors, \emph{{Field dynamics and tunneling in a flux
  landscape}}, \href{http://dx.doi.org/10.1103/PhysRevD.78.083534}{\emph{Phys.
  Rev.} {\bf D78} (2008) 083534}, [\href{https://arxiv.org/abs/0805.3705}{{\tt
  0805.3705}}].

\bibitem{1011.6588}
P.~Ahlqvist, B.~R. Greene, D.~Kagan, E.~A. Lim, S.~Sarangi and I.-S. Yang,
  \emph{{Conifolds and Tunneling in the String Landscape}},
  \href{http://dx.doi.org/10.1007/JHEP03(2011)119}{\emph{JHEP} {\bf 03} (2011)
  119}, [\href{https://arxiv.org/abs/1011.6588}{{\tt 1011.6588}}].

\bibitem{1108.1394}
A.~P. Braun, N.~Johansson, M.~Larfors and N.-O. Walliser, \emph{{Restrictions
  on infinite sequences of type IIB vacua}},
  \href{http://dx.doi.org/10.1007/JHEP10(2011)091}{\emph{JHEP} {\bf 10} (2011)
  091}, [\href{https://arxiv.org/abs/1108.1394}{{\tt 1108.1394}}].

\bibitem{Brown:2017osf}
A.~R. Brown, \emph{{Hyperinflation}},
  \href{https://arxiv.org/abs/1705.03023}{{\tt 1705.03023}}.

\bibitem{Mizuno:2017idt}
S.~Mizuno and S.~Mukohyama, \emph{{Primordial perturbations from inflation with
  a hyperbolic field-space}},  \href{https://arxiv.org/abs/1707.05125}{{\tt
  1707.05125}}.

\bibitem{Achucarro:2016fby}
A.~Achúcarro, V.~Atal, C.~Germani and G.~A. Palma, \emph{{Cumulative effects
  in inflation with ultra-light entropy modes}},
  \href{http://dx.doi.org/10.1088/1475-7516/2017/02/013}{\emph{JCAP} {\bf 1702}
  (2017) 013}, [\href{https://arxiv.org/abs/1607.08609}{{\tt 1607.08609}}].

\bibitem{Achucarro:2017ing}
A.~Achúcarro, R.~Kallosh, A.~Linde, D.-G. Wang and Y.~Welling,
  \emph{{Universality of multi-field $\alpha$-attractors}},
  \href{https://arxiv.org/abs/1711.09478}{{\tt 1711.09478}}.

\bibitem{Kobayashi:2010fm}
T.~Kobayashi and S.~Mukohyama, \emph{{Effects of Light Fields During
  Inflation}}, \href{http://dx.doi.org/10.1103/PhysRevD.81.103504}{\emph{Phys.
  Rev.} {\bf D81} (2010) 103504}, [\href{https://arxiv.org/abs/1003.0076}{{\tt
  1003.0076}}].

\bibitem{Cremonini:2010sv}
S.~Cremonini, Z.~Lalak and K.~Turzynski, \emph{{On Non-Canonical Kinetic Terms
  and the Tilt of the Power Spectrum}},
  \href{http://dx.doi.org/10.1103/PhysRevD.82.047301}{\emph{Phys. Rev.} {\bf
  D82} (2010) 047301}, [\href{https://arxiv.org/abs/1005.4347}{{\tt
  1005.4347}}].

\bibitem{Cremonini:2010ua}
S.~Cremonini, Z.~Lalak and K.~Turzynski, \emph{{Strongly Coupled Perturbations
  in Two-Field Inflationary Models}},
  \href{http://dx.doi.org/10.1088/1475-7516/2011/03/016}{\emph{JCAP} {\bf 1103}
  (2011) 016}, [\href{https://arxiv.org/abs/1010.3021}{{\tt 1010.3021}}].

\bibitem{vandeBruck:2014ata}
C.~van~de Bruck and M.~Robinson, \emph{{Power Spectra beyond the Slow Roll
  Approximation in Theories with Non-Canonical Kinetic Terms}},
  \href{http://dx.doi.org/10.1088/1475-7516/2014/08/024}{\emph{JCAP} {\bf 1408}
  (2014) 024}, [\href{https://arxiv.org/abs/1404.7806}{{\tt 1404.7806}}].

\bibitem{Turzynski:2014tza}
S.~Renaux-Petel and K.~Turzynski, \emph{{On reaching the adiabatic limit in
  multi-field inflation}},
  \href{http://dx.doi.org/10.1088/1475-7516/2015/06/010}{\emph{JCAP} {\bf 1506}
  (2015) 010}, [\href{https://arxiv.org/abs/1405.6195}{{\tt 1405.6195}}].

\bibitem{0201028}
S.~Kachru, M.~B. Schulz and S.~Trivedi, \emph{{Moduli stabilization from fluxes
  in a simple IIB orientifold}},
  \href{http://dx.doi.org/10.1088/1126-6708/2003/10/007}{\emph{JHEP} {\bf 10}
  (2003) 007}, [\href{https://arxiv.org/abs/hep-th/0201028}{{\tt
  hep-th/0201028}}].

\bibitem{0506179}
J.~Gomis, F.~Marchesano and D.~Mateos, \emph{{An Open string landscape}},
  \href{http://dx.doi.org/10.1088/1126-6708/2005/11/021}{\emph{JHEP} {\bf 11}
  (2005) 021}, [\href{https://arxiv.org/abs/hep-th/0506179}{{\tt
  hep-th/0506179}}].

\bibitem{1405.0283}
M.~Arends, A.~Hebecker, K.~Heimpel, S.~C. Kraus, D.~Lust, C.~Mayrhofer et~al.,
  \emph{{D7-Brane Moduli Space in Axion Monodromy and Fluxbrane Inflation}},
  \href{http://dx.doi.org/10.1002/prop.201400045}{\emph{Fortsch. Phys.} {\bf
  62} (2014) 647--702}, [\href{https://arxiv.org/abs/1405.0283}{{\tt
  1405.0283}}].

\bibitem{1703.09729}
A.~Landete, F.~Marchesano, G.~Shiu and G.~Zoccarato, \emph{{Flux Flattening in
  Axion Monodromy Inflation}},
  \href{http://dx.doi.org/10.1007/JHEP06(2017)071}{\emph{JHEP} {\bf 06} (2017)
  071}, [\href{https://arxiv.org/abs/1703.09729}{{\tt 1703.09729}}].

\bibitem{Verrill:2001}
H.~A. Verrill, \emph{{Algorithm for Drawing Fundamental Domains}},
  {\emph{http://wstein.org/Tables/fundomain/index2.html} (2001) }.

\bibitem{Stein:2003}
W.~A. Stein, \emph{{lecture notes on `Modular Abelian Varieties'}},
  {\emph{http://wstein.org/edu/Fall2003/252/lectures/index.html} (2003) }.

\bibitem{Blumenhagen:2014nba}
R.~Blumenhagen, D.~Herschmann and E.~Plauschinn, \emph{{The Challenge of
  Realizing F-term Axion Monodromy Inflation in String Theory}},
  \href{http://dx.doi.org/10.1007/JHEP01(2015)007}{\emph{JHEP} {\bf 01} (2015)
  007}, [\href{https://arxiv.org/abs/1409.7075}{{\tt 1409.7075}}].

\bibitem{Garcia-Etxebarria:2014wla}
I.~García-Etxebarria, T.~W. Grimm and I.~Valenzuela, \emph{{Special Points of
  Inflation in Flux Compactifications}},
  \href{http://dx.doi.org/10.1016/j.nuclphysb.2015.08.008}{\emph{Nucl. Phys.}
  {\bf B899} (2015) 414--443}, [\href{https://arxiv.org/abs/1412.5537}{{\tt
  1412.5537}}].

\bibitem{Bizet:2016paj}
N.~Cabo~Bizet, O.~Loaiza-Brito and I.~Zavala, \emph{{Mirror quintic vacua:
  hierarchies and inflation}},
  \href{http://dx.doi.org/10.1007/JHEP10(2016)082}{\emph{JHEP} {\bf 10} (2016)
  082}, [\href{https://arxiv.org/abs/1605.03974}{{\tt 1605.03974}}].

\bibitem{Donagi:2017mhd}
R.~Donagi and E.~Sharpe, \emph{{On the global moduli of Calabi-Yau
  threefolds}},  \href{https://arxiv.org/abs/1707.05322}{{\tt 1707.05322}}.

\bibitem{1604.05325}
A.~Gerhardus and H.~Jockers, \emph{{Quantum periods of Calabi–Yau
  fourfolds}},
  \href{http://dx.doi.org/10.1016/j.nuclphysb.2016.09.021}{\emph{Nucl. Phys.}
  {\bf B913} (2016) 425--474}, [\href{https://arxiv.org/abs/1604.05325}{{\tt
  1604.05325}}].

\bibitem{Abe:2014xja}
H.~Abe, T.~Kobayashi and H.~Otsuka, \emph{{Natural inflation with and without
  modulations in type IIB string theory}},
  \href{http://dx.doi.org/10.1007/JHEP04(2015)160}{\emph{JHEP} {\bf 04} (2015)
  160}, [\href{https://arxiv.org/abs/1411.4768}{{\tt 1411.4768}}].

\end{thebibliography}\endgroup
\bibliographystyle{JHEP}
\end{document}